%% file: main.tex
\documentclass{article}
\usepackage[dvipsnames]{xcolor}
\definecolor{Gray}{gray}{0.9}
\usepackage{microtype}
\usepackage{graphicx}
\usepackage{subfigure}
\usepackage{booktabs} 
\usepackage{authblk} 
\usepackage{hyperref} 
\usepackage{pifont} 
\usepackage{booktabs} 
\newcommand{\cmark}{\ding{51}} 
\newcommand{\xmark}{\ding{55}} 
\pdfoutput=1
\usepackage{tikz}

\usepackage[accepted]{icml2024}

\usepackage{amsmath}
\usepackage{amssymb}
\usepackage{mathtools}
\usepackage{amsthm}

\usepackage[capitalize,noabbrev]{cleveref}

\theoremstyle{plain}    
\newtheorem{theorem}{Theorem}[section]

\newtheorem{lemma}[theorem]{Lemma}

\theoremstyle{definition}
\newtheorem{definition}[theorem]{Definition}

\theoremstyle{remark}

\usepackage[textsize=tiny]{todonotes}

\usepackage{etoc}
\usepackage{framed} 
\usepackage{longtable}
\usepackage{etoc}
\etocdepthtag.toc{mtchapter}
\etocsettagdepth{mtchapter}{subsection}
\etocsettagdepth{mtappendix}{none}

\usepackage{framed} 
\definecolor{shadecolor}{rgb}{0.94, 0.97, 1.0}

\usepackage{url}

\usepackage[resetlabels]{multibib}
\usepackage{multirow}
\usepackage{multicol}
\usepackage{enumitem}
\usepackage[normalem]{ulem}

\usepackage{threeparttable}
\usepackage{adjustbox}
\usepackage{wrapfig}
\usepackage{nomencl} 

\makenomenclature

\definecolor{cite_color}{HTML}{114083}
\definecolor{link_color}{RGB}{153, 0,0}  
\definecolor{url_color}{RGB}{153, 102,  0}
\definecolor{emp_color}{RGB}{0,0,255}
\definecolor{aliceblue}{rgb}{0.94, 0.97, 1.0}
\hypersetup{
 colorlinks,
 citecolor=cite_color,
 linkcolor=link_color,
 urlcolor=url_color}

\usepackage{caption}
\usepackage{pifont}
\usepackage{makecell}
\usepackage{subfigure} 
\usepackage{amsmath}
\usepackage{multirow}
\usepackage{multicol}
\usepackage{enumitem}
\usepackage{ulem}
\usepackage{threeparttable}
\usepackage{wrapfig}
\usepackage{scrextend}
\usepackage{thm-restate}
\usepackage{array,hhline}
\usepackage{amsmath,amsfonts}
\usepackage{tabularborder}
\usepackage[ruled]{algorithm2e}
\usepackage{setspace}
\usepackage{mathtools}
\DeclarePairedDelimiterX{\infdivx}[2]{(}{)}{%
  #1\;\delimsize\|\;#2%
}
\usepackage{cancel}

\definecolor{mypink}{RGB}{255,211,150}

\SetKwComment{Comment}{$\triangleright$\ }{}
\usepackage{amssymb}
\usepackage{xspace}
\usepackage{wrapfig,tikz}
\usepackage{siunitx}

\usepackage{cleveref}

\usepackage{lmodern}
\usepackage{anyfontsize}
\usepackage[ruled]{algorithm2e}

\input{icml_math_commands}

\icmltitlerunning{From thermodynamics to protein design: Diffusion models for biomolecule generation towards autonomous protein engineering}

\begin{document}

\twocolumn[
\icmltitle{From thermodynamics to protein design: Diffusion models for biomolecule generation towards autonomous protein engineering}

\begin{icmlauthorlist}
\icmlauthor{Wenran LI}{a}
\icmlauthor{Xavier F. Cadet}{b}
\icmlauthor{David Medina-Ortiz}{c,d}
\icmlauthor{Mehdi D. Davari}{e}
\icmlauthor{Ramanathan Sowdhamini}{f,g,h}
\icmlauthor{Cedric Damour}{i}
\icmlauthor{Yu Li}{j}
\icmlauthor{Alain Miranville}{k}
\icmlauthor{Frederic Cadet}{a,l}
\end{icmlauthorlist}

\icmlaffiliation{a}{University of Paris City \& University of Reunion, INSERM, BIGR, DSIMB, F-75015 Paris, France}
\icmlaffiliation{b}{Department of Computing, Imperial College London, United Kingdom}
\icmlaffiliation{c}{Department of Computer Engineering, University of Magallanes, Punta Arenas, Chile}
\icmlaffiliation{d}{Centre for Biotechnology and Bioengineering, Universidad de Chile, Santiago, Chile}
\icmlaffiliation{e}{Department of Bioorganic Chemistry, Leibniz Institute of Plant Biochemistry, Halle 06120, Germany}
\icmlaffiliation{f}{National Centre for Biological Science, TIFR, Bangalore, India}
\icmlaffiliation{g}{Molecular Biophysics Unit, Indian Institute of Science, Bangalore, India}
\icmlaffiliation{h}{Institute of Bioinformatics and Applied Biotechnology, Bangalore, India}
\icmlaffiliation{i}{EnergyLab, EA 4079, Faculty of Sciences and Technology, University of Reunion, France}
\icmlaffiliation{j}{School of Information Science and Technology and Beijing Institute of Artificial Intelligence, China}
\icmlaffiliation{k}{Laboratoire de Mathématiques Appliquées, University Le Havre Normandie, le Havre, France}
\icmlaffiliation{l}{PEACCEL, Artificial Intelligence Department, AI for Biologics, 75013, Paris, France}
\icmlcorrespondingauthor{Frederic Cadet}{\texttt{frederic.cadet.run@gmail.com}}

\vskip 0.3in
]

\printAffiliationsAndNotice{}  

\begin{abstract}
Protein design with desirable properties has been a significant challenge for many decades. Generative artificial intelligence is a promising approach and has achieved great success in various protein generation tasks. Notably, diffusion models stand out for their robust mathematical foundations and impressive generative capabilities, offering unique advantages in certain applications such as protein design. In this review, we first give the definition and characteristics of diffusion models and then focus on two strategies: Denoising Diffusion Probabilistic Models (DDPM) and Score-based Generative Models (SGM), where DDPM is the discrete form of SGM. Furthermore, we discuss their applications in protein design, peptide generation, drug discovery, and protein-ligand interaction. Finally, we outline the future perspectives of diffusion models to advance autonomous protein design and engineering. The $E(3)$ group consists of all rotations, reflections, and translations in three-dimensions. The equivariance in the $E(3)$ group can maintain the physical stability of the $N-C_{\alpha}-C$ frame of each amino acid as much as possible, and we reflect on how to keep the diffusion model $E(3)$ equivariant for protein generation.
\vspace{1em}
\end{abstract}

\addtocontents{toc}{\protect\setcounter{tocdepth}{-1}}
\section{Introduction}

For decades, protein engineering and protein design tasks have been regarded as NP-hard optimization problems,the algorithmic challenges continue to persist despite advancements in computational methods. \cite{mukhopadhyay_BRIEF2014,pierce_Protein2002}. As the number of residues increases from 75 to 200, the number of conformations increases from $O(n^{75})$ to $O(n^{200})$, where $n$ is the average number of rotamers per position. Researchers have been working to explore effective methods to bridge the sizeable gap. Due to their ability to learn complex patterns for large datasets, deep learning approaches have been applied to various tasks such as protein structure prediction, sequence design for specific functions, and \textit{de novo} protein design (DNPD) \cite{watsonNovoDesignProtein2023}. Generative modeling is a subfield of ML that focuses on developing algorithms capable of generating new data samples that resemble the data distribution from a given training dataset. Successful applications of generative modeling have highlighted the potential of protein design by modeling the probability distribution of protein sequences. Techniques such as variational autoencoders (VAE) and generative adversarial networks (GAN) have been employed on generation problems for protein sequences and structures \cite{rossetto_GANDALF2019,tucs_Generating2020}. Alternatively, diffusion models have given amazing results for image, audio, and text synthesis, while being relatively simple to implement.
Diffusion models are related to stochastic differential equations (SDEs), making their theoretical properties particularly intriguing. These models have shown significant advantages in modeling complex distributions and have thus gained traction in protein engineering \cite{tang2024survey}. Using their mathematical foundations, diffusion models offer a promising framework for addressing challenges in protein design.

Building on these foundations, a diffusion probabilistic model \cite{kloeden1992stochastic} uses a parameterized Markov chain trained by variational inference. This approach enables the generation of samples that align with the data distribution within finite time, providing a structured and efficient mechanism for generative tasks.
Transitions of this chain are learned to reverse a diffusion process, which is a Markov chain that gradually adds noise to the data in the opposite direction of sampling until signal is destroyed. Diffusion models address key challenges faced by other generative approaches: they overcome the difficulty of accurately matching posterior distributions in VAEs, mitigate the instability arising from the adversarial training objectives in GANs, and excel in protein generation tasks, particularly in producing structures with improved atom stability\cite{chen_Overview2024,tang_Survey2024,li_Generalization2024}. \par
 The concept of equivariance \cite{batzner_Equivariant2022} arises naturally in machine learning of atomistic systems: physical properties have well-defined transformation properties under translation, reflection, and rotation of a set of atoms. Several reviews on the application of diffusion modeling to the generation of biomolecules have been published \cite{norton2024sifting,guoDiffusionModelsBioinformatics2023,zhang_Survey2023,goles_Peptidebased2024}; they have surveyed some diffusion models that can address various bioinformatics problems, such as denoising cryo-EM data, single-cell gene expression analysis, and protein design (for details, see Appendix. I.). However, most reviews have not discussed the common mathematical features and the importance of equivariance properties. 
 \par
 The motivation for this work is to provide advanced and comprehensive insights into the development, evaluation, and comparison of diffusion models, explaining the advantages and disadvantages of these approaches compared to other generative models, and the future directions and perspectives of diffusion models to assist the protein design. \par

The main \textbf{contributions} of this review include:
\begin{enumerate}
\item[$\bullet$] An accessible introduction to the fundamentals of diffusion models and equivariance.
\item[$\bullet$] A fairly detailed overview of the applications of 56 diffusion models in biomolecule design (for more details, see Appendix. A.).
\item[$\bullet$] A discussion on the future development of diffusion models to assist in biomolecule design.
\end{enumerate}
This work explores the generation of different biomolecules through diffusion models, emphasizing protein design.

\section{Theoretical preparation}

This section introduces two common diffusion models, DDPM and SGM, to lay the foundation for the following sections. In addition, we give the concepts of symmetry and equivariance and the relationship between them. The relationship between the molecular structure and the model is also revealed.
\subsection{Diffusion models}

\begin{figure}[ht]
    \centering
    \vspace{0.6cm}
    \includegraphics[width=0.35\textwidth, bb=80 80 560 108]{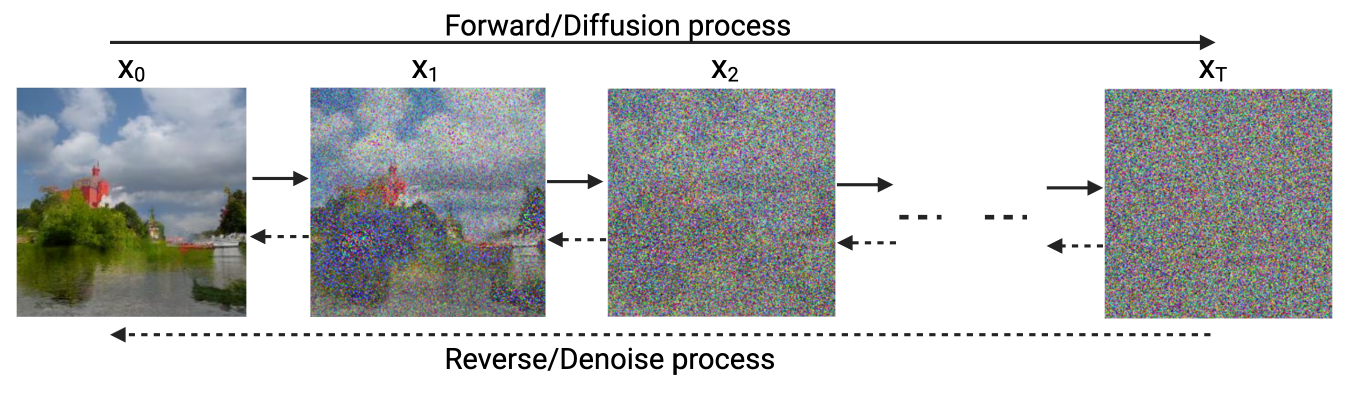}
    \vspace{0.6cm}
    \caption{Visualization of diffusion models operating on the image generation. During the diffusion process, the image becomes blurred until it becomes a Gaussian distribution. The reverse process is a denoising process, and the image gradually becomes clear.}
    \label{illustration_diffusion}
\end{figure}

A diffusion model is a deep generative model based on two stages: a forward diffusion stage and a reverse diffusion stage. In the forward diffusion stage, the input data are gradually perturbed over several steps by adding Gaussian noise. In the reverse phase, a model restores the original input data by learning to reverse the diffusion process step by step. Figure \ref{illustration_diffusion} illustrates how a diffusion model works to generate an image.
\par

In the discrete form, for a sufficiently large time $T>0$, $t=0,1,..., T$, with the random variable $x_{0}\in\mathbb{R}^{n}$, where $n$ is the dimension, the forward process iteratively adds isotropic Gaussian noise to the sample. The Gaussian transition kernel is set as:
\begin{align}
q(x_{t}|x_{t-1})&=\mathcal{N}(\sqrt{1-\beta_{t}}x_{t-1}, \beta_{t}I),\\
q(x_{1:T}|x_{0})&=\prod_{t=1}^{T}q(x_{t}|x_{t-1}),
\label{forward}
\end{align}
where the $\beta_{t}$ are chosen according to a fixed variance scheme \cite{song_Denoising2022,croitoru_Diffusion2023,rombach_HighResolution2022b}. Noisy data $x_{t}$ can be sampled directly from $x_{0}$:
\begin{equation}
\label{noisy_data}
x_{t}=\sqrt{\alpha_{t}}x_{0}+\sqrt{1-\alpha_{t}}\epsilon,
\end{equation}
where $\epsilon \sim \mathcal{N}(0,I)$ and $\alpha_{t}=\prod_{s=1}^{t}(1-\beta_{s}).$

\par
While the reverse process, starting from noise $x_{T}\sim \mathcal{N}(0,I)$, aims to learn the process of denoising:
\begin{align}
p_{\theta}(x_{0})&=p(x_{T})\prod_{t=1}^{T}p_{\theta}(x_{t-1}|x_{t}); \\
p_{\theta}(x_{t-1}|x_{t})&=\mathcal{N}(x_{t-1};\mu_{\theta}(x_{t},t),\sigma_{\theta}(x_{t},t)),
\end{align}
i.e. to learn $p_{\theta}(x_{t-1}|x_{t})$ using a model with hyperparameters $\theta$. Here
\[
\mu_{\theta}(x_{t}, t)=\frac{1}{\sqrt{1-\beta_{t}}}(x_{t}-\frac{\beta_{t}}{\sqrt{1-\alpha_{t}}}\sigma_{\theta}(x_{t},t)),
\]
the DDPM aims to approximate $\epsilon$ using a parametric model structured as $\sigma_{\theta}$. The objective function can be written as follows:
\[
\theta^{*}=\mathop{\arg\min}_{\theta} \mathbb{E}_{x_{0},t,\epsilon}[\|\epsilon-\sigma_{\theta}(\sqrt{\alpha_{t}}x_{0}+\sqrt{1-\alpha_{t}}\epsilon,t)\|^{2}].
\]

In the continuous form \cite{kingma_Variational2023}, the following stochastic differential equation (SDE) \cite{kloeden1992stochastic} has the same transition distribution $q(x_{t}|x_{0})$ as in equation \eqref{forward} for any $t\in[0,T]$:
\[
dx=f(t)x_{t}dt+g(t)d\omega_{t},
\]
where $\omega_{t}$ is the standard Wiener process, $f(t)$ is a drift term that typically describes a time-dependent scaling of the data,  and $g(t)$ is a scalar function known as the diffusion coefficient. \par
\cite{song2020score} indicated that the following time reversal SDE and probability flow ordinary differential equation (ODE) preserve the marginal distribution for $x_{T}\sim p_{\theta}(x_{T})$:
\begin{align}
dx&=[f(x,t)-g(t)^{2}\nabla_{x}\log p_{\theta}(x)]dt+g(t)d\bar{\omega}_{t}, \\
dx&=[f(x,t)-\frac{1}{2}g(t)^{2}\nabla_{x}\log p_{\theta}(x)]dt
\end{align}
where $\bar{\omega}_{t}$ is the reverse Wiener process, $\nabla_{x}\log p_{\theta}(x)$ is the Stein score.
Score-based generative models learn the gradient of the probability distribution rather than the distribution itself, i.e,
\[
\theta^{*}= \mathop{\arg\min}_{\theta} \mathbb{E}_{x_{0},t,\epsilon}[\|s_{t,\theta}(x_{t})-\nabla_{x_{t}}\log p_{\theta}(x_{t}|x_{0})\|^{2}].
\]
Further descriptions about diffusion models are provided in Appendix. B.

    \subsection{Geometric symmetry and equivariance}
Geometric symmetry and equivariance are related concepts in mathematics and machine learning, especially when dealing with transformations like rotations, translations, and reflections. 

\begin{definition}
    (Symmetry) \cite{cohen2021equivariant} Let $X$ denote the input space, $Y$ the label space, and $w$ the weight space, let $f: X\times W \rightarrow Y$ denote a model. A transformation $g: W\rightarrow W$ is a symmetry of the parameterization if
\[
f(x,gw)=f(x,w)\qquad \text{for all } x\in X\text{ and }w\in W
\]
\end{definition}

\begin{definition}
        (Equivariant) \cite{bronstein2021geometric} Let $\rho_{g}: X \rightarrow X$ be a set of transformations on $X$ for the abstract group $g\in G$. We say a function $f: X\rightarrow Y$ is equivariant to $g$ if there exists an equivalent transformation on its output space $\rho '_{g}: Y\rightarrow Y$ such that:
\[
f(\rho_{g}(x))=\rho'_(f(x)).
\]
\end{definition}
Symmetry typically refers to the static properties of shapes, patterns, or systems. It is used to describe the geometric conformation in proteins. Equivariance, on the other hand, refers to the dynamic relationships between input and output under transformations. Most of the models discussed in this paper are equivariant models.

\section{Diffusion model for protein generation}

This section discusses the generation of protein sequence and structure separately.

\subsection{Sequence Generation}
\label{protein_sequence}

Sequence generation models usually regard amino acids as the word, input them to language models for feature extraction first, then input them to diffusion models for generation. \par

    \textbf{TaxDiff} \cite{zongyingTaxDiffTaxonomicGuidedDiffusion2024} combines the denoise transformer with the diffusion model to learn taxonomically guided over the space of protein sequences and thus fulfills the requirements of downstream tasks in biology. \textbf{EvoDiff} \cite{alamdariProteinGenerationEvolutionary2023} presents order-agnostic autoregressive diffusion models (DAOMs) and discrete denoising diffusion probabilistic models (D3PM) to generate highly realistic, diverse and structurally plausible proteins.  

\par
\textbf{DPLM} \cite{wang_Diffusion2024} initially trained with masked language models (MLMs), then continuously trained with the diffusion objective, demonstrates a strong generative capability for protein sequences. 
DPLM-2 \cite{wang2024dplm} is a multimodal protein foundation model that extends DPLM to accommodate both sequences and structures, where foundation models are large deep learning neural networks that have changed the way data scientists approach machine learning. 
For assessing the feasibility of the sequences, \cite{zongyingTaxDiffTaxonomicGuidedDiffusion2024} used OmegaFold \cite{wu2022high} to predict their corresponding structures and calculate the average predicted Local Distance Difference Test (pLDDT) across the entire structure, which reflects OmegaFold’s confidence in its structure prediction for each residue on sequences level. We compare the pLDDT of the models mentioned above in Table \ref{plddt}. We can see from Table \ref{plddt} that the pLDDT score of the sequences sampled by DPLM-2 is close to that of DPLM. This score suggests that DPLM-2 largely retains its sequence generation capability inherited from sequence pre-training in DPLM. 

\begin{table}[!htpb]
\caption{pLDDT results of the diffusion models: EvoDiff, TasDiff, DPLM and DPLM-2. DPLM achieves the best feasibility among them.}
\label{plddt}
\begin{tabular}{|l|l|l|l|l|}
\hline
Model             & EvoDiff & TaxDiff & DPLM  & DPLM-2 \\ \hline
pLDDT($\uparrow$) & 44.29   & 68.89   & 83.25 & 82.25  \\ \hline
\end{tabular}
\end{table}
Evolutionary scale modelling (ESM) \cite{lin2023evolutionary} is a class of language models applied to the generation of protein sequences. 
\textbf{ForceGen} \cite{niForceGenEndtoendNovo2024} develops a pLDM by combining the ESM Metagenomic Atlas \cite{lin2023evolutionary}, a model of the ESM family, with an attention-based diffusion model \cite{ni_Generative2023} to generate a protein sequence and structure with non-mechanical properties. \par

\subsection{Structure Generation}
\label{protein_strucutre}

Generating a backbone is a difficult task because a backbone should fulfill the following three criteria:
\begin{itemize}
\item \textbf{Physically realizable:} We can find the sequence that folds into the generated structure \cite{martin2008long}.
\item \textbf{Functional:} We aim for conditional sampling under diverse functional constraints without retraining \cite{mandell2009backbone}.
\item \textbf{Generalizability:} We hope that the model has multiple application scenarios\cite{murphy2012increasing}.
\end{itemize} 

For the above criteria, we introduce some models that in our opinion best meet the standards in order, and discuss the effects of these models.
\par

\subsubsection{Physically realizable model: Diffusion on $SE(3)$ group}
\label{protein_se3}
$SE(3)$ is the notation for the special Euclidean 3D group that includes translational and rotational isometric transformations and keeps the volume constant (see more details in Appendix. D). This mathematical framework is particularly relevant for modeling molecular systems, where maintaining spatial invariance is crucial for accurate predictions.
\par
Building on this principle, \textbf{RFDiffusion} \cite{watsonNovoDesignProtein2023} repurposes RoseTTAFold \cite{baek_Accurate2021} to perform reverse diffusion. The $SE(3)$-equivariance of RoseTTAFold underpins RFDiffusion's ability to respect these isometric transformations during the generative process. RFDiffusion has also been effectively applied in the design of peptide binders \cite{vazqueztorresNovoDesignHighaffinity2024,liu_Diffusing2024}. By fine-tuning RoseTTAFold All-Atom (RFAA) \cite{krishna2024generalized}, a neural network for predicting biomolecular structures, to diffusion denoising tasks, \textbf{RFDiffusionAA} generates folded protein structures surrounding the small molecule from random residue distributions.
\textbf{ProteinGenerator} \cite{lisanza2023joint} is a sequence space diffusion model based on RoseTTAFold that simultaneously generates protein sequences and structures. The success rate of ProteinGenerator in generating long sequences that fold to the designed structure is lower than RFDiffusion, this may reflect the intrinsic difference between diffusion in sequence and structure spaces.\par
\textbf{FrameDiff} \cite{yimSEDiffusionModel2023} is a diffusion model in the Lie group \cite{watson2022broadly} $SE(3)_{0}^{N}$ for the generation of protein backbones. It's forward process is,
    \[
    dT^{(t)} =[0, -\frac{1}{2}PX^{(t)}]dt+[dB_{so(3)^{N}}^{t}, PdB_{\mathbb{R}^{3N}}^{(t)}],
    \]
where $P\in \mathbb{R}^{3N\times 3N}$ is the projection matrix removing the center of mass $\frac{1}{N}\sum_{n=1}^{N}x_{n}$, and $T^{(t)}_{t\ge 0}=(R^{(t)},X^{(t)})_{t\ge 0}$ is a stochastic process on $SE(3)_{0}^{N}$ with invariant measure $\mathcal{N}(0,\mathbf{I}d)^{\otimes N}\otimes \mathcal{U}(SO(3))^{\otimes N}$ pushforward by $P$. The backward process $(\overleftarrow{T}^{t})_{t\in [0,T_{F}]}=([\overleftarrow{R}^{t},\overleftarrow{X}^{(t)}])_{t\in [0,T_{F}]}$ is given by 
\begin{align}
    d\overleftarrow{R}^{(t)}&=\nabla_{r}\log p_{T_{F-t}}(\overleftarrow{T}^{(t)})dt+d\mathbf{B}_{SO(3)^{N}}^{(t)},  \\
    d\overleftarrow{X}^{(t)}&=P\{\frac{1}{2}\overleftarrow{X}^{(t)}+\nabla_{x}\log p_{T_{F-t}}(\overleftarrow{T}^{(t)})\}dt+Pd\mathbf{B}_{\mathbb{R}^{3N}}^{(t)}. 
\end{align}
This model applies Invariant Point Attention (IPA) \cite{jumper_Highly2021} to keep the updates of residues in coordinate space that are $SE(3)$-invariant. 
\par
FrameDiff has been used for inpainting protein structures and motif scaffolding, named \textbf{FrameDiPT} \cite{zhangFrameDiPTSEDiffusion2023} and \textbf{TDS} \cite{wu2024practical}, respectively.
\textbf{VFN-Diff} \cite{mao_Novo2023} replaces the IPA in FrameDiff with Vector Field Networks (VFN), which is also the $SE(3)$ equivariant model. VFN-Diff significantly outperforms FrameDiff in terms of design capability and diversity.
\par

\textbf{Genie} \cite{lin_Generating2023} combine aspects of the $SE(3)$-equivariant reasoning machinery of IPA with DDPMs to create a $SE(3)$-equivariant denoiser $\epsilon_{\theta}(F(x_{t}),t)$ in the protein generation process. \textbf{Genie2} \cite{lin_Out2024} extended Genie to motif scaffolding, and introduced a novel multi-motif framework that designs co-occurring motifs without needing to specify inter-motif positions and orientations in advance. \par

\begin{figure*}[ht]
\centering
\includegraphics[bb=0 0 790 520, width=\textwidth]{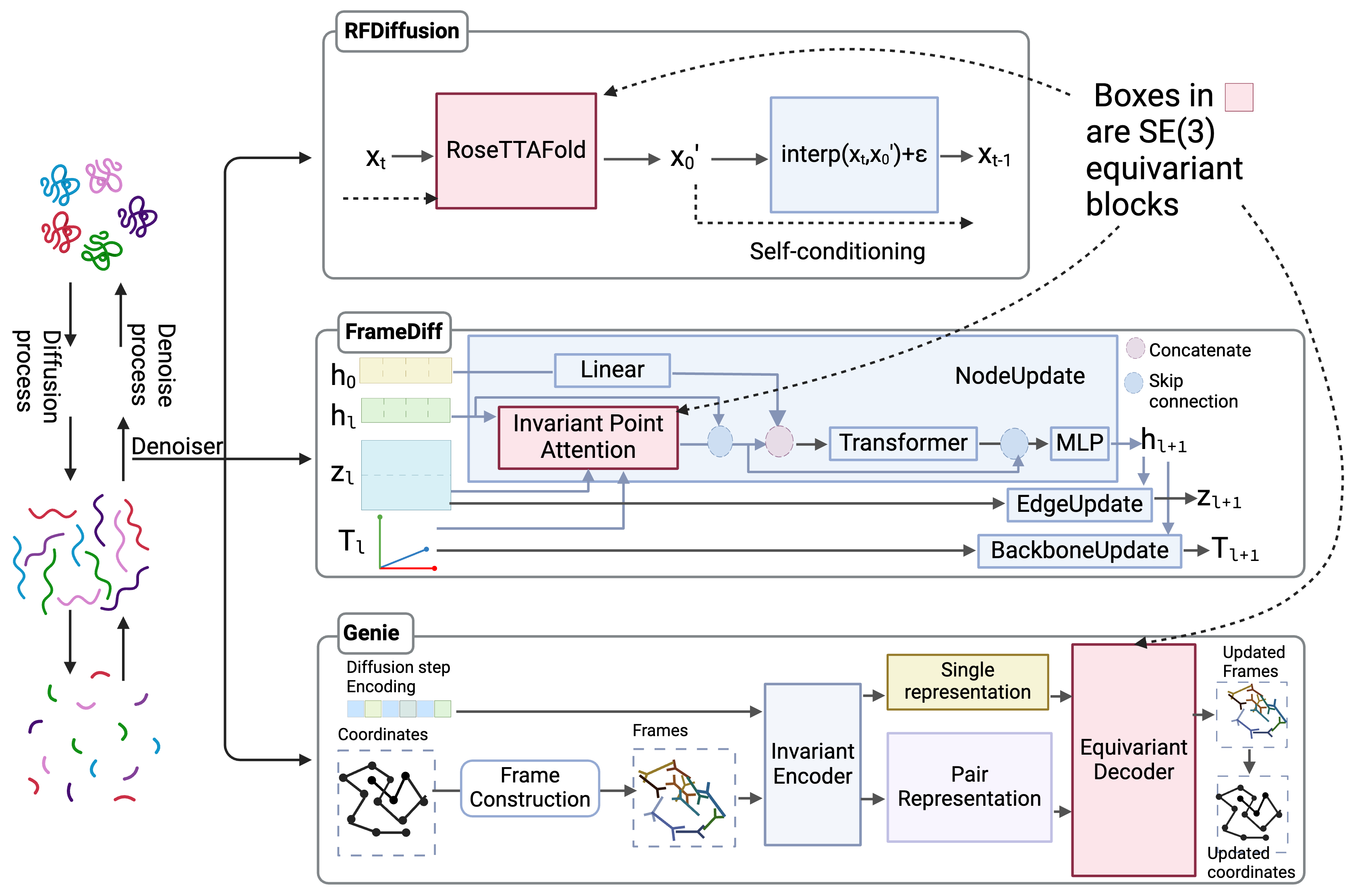}
\caption{$SE(3)$ equivariant diffusion models for protein structure generation. RFDiffusion, FrameDiff and Genie utilize RoseTTAFold, IPA and $SE(3)$-equivariant denoiser as the single step of the denoise process in the diffusion model, respectively. Boxes in pink color are $SE(3)$ equivariant blocks. $SE(3)$ equivariant keeps the frames of each amino acid physically stable.}
\label{RFDiffusion}
\end{figure*}

Figure \ref{RFDiffusion} shows that RFDiffusion, FrameDiff and Genie all utilize $SE(3)$ equivariant natural network into the denoiser. This kind of architecture will keep the $N-C_{\alpha}-C$ frame of each amino acid residue invariant to global rotations and translations.
As special subsets of $SE(3)$ equivariant models, some protein generation models such as \textbf{ProtDiff-SMCDiff} \cite{trippe_Diffusion2023} satisfy $E(3)$ equivariance. They can additionally keep consistency for permutation and translation. This kind of model is highly interesting in molecular design, but until now few protein generation models satisfy this $E(3)$ equivariance property (See more details in Appendix. E).

\subsubsection{Model with strong functionality}
\label{functional}

Protein design projects often involve complex and composite requirements that vary over time. \textbf{Chroma} \cite{ingrahamIlluminatingProteinSpace2023} explores a programmable generative process with custom energy functions, which aims to make the generated protein have desired properties and functions, such as symmetry, substructure, shape and semantics.\par
Table \ref{tab:survey_comparison} shows the comparison of several classical models with their advantages, disadvantages and performances.

\begin{table*}[t]
    \centering
    \caption{Comparison of different protein structure models: Advantages, disadvantages, and performances on 100 amino acid proteins.}
    \label{tab:survey_comparison}
    \footnotesize
    \scalebox{0.95}{
    \begin{tabular}{p{2cm}|p{5cm}|p{3cm}|p{5.5cm}}
\hline
\textbf{Models} &
  \textbf{Strength} &
  \textbf{Weakness} &
  \textbf{Potential application areas} \\ \hline
ProtDiff-SMCDiff &
  Computational efficiency &
  Complexity &
  Peptide generation; motif scaffolding \\ 
AlphaFold3 &
  Adaptability to different biomolecule types &
  Hallucinations in disordered regions &
  Dynamical behavior of biomolecular systems \\ 
RFDiffusion &
  High accuracy; good at conditional tasks &
  Low flexibility &
  Protein-ligand interaction \\ 
FrameDiff &
  Theoretical; does not require pre-trained structure predictors &
  Complexity &
  Peptide generation \\ 
Genie &
  Simplicity; designability, diversity &
  Capacity limited &
  Longer protein design \\ 
Chroma &
  Programmability; jointly models structures and sequences &
  Complexity &
  Peptide generation \\ \hline
\end{tabular}
}
\end{table*}

\begin{figure*}[t]
    \centering
    \begin{tikzpicture}
        \draw[->] (-0.1,0) -- (14.5,0);

        \foreach \x in {0,1.8,3.2,4.5,5.8,7,8.2, 9.5, 11,12.5, 14}
            \draw (\x cm,3pt) -- (\x cm,-3pt);

        \draw (0,0) node[below=3pt] {2022.03}
                  node[above=5pt, text=red] {EDM}; 
        \draw (1.8,0) node[below=3pt] {22.10}
                  node[above=5pt, text=red] {DiffSBDD}
                  node[above=20pt, text=orange] {RFDiffusion}; 
        \draw (3.2,0) node[below=3pt] {2023.01}
                  node[below=15pt, text=blue] {Genie}; 
        \draw (4.5,0) node[below=3pt] {23.02}
                  node[above=5pt, text=red] {MiDi}
                  node[above=20pt, text=blue] {FrameDiff};
        \draw (5.8,0) node[below=3pt] {23.05}
                  node[below=15pt, text=orange] {ProteinGenerator}; 
        \draw (7,0) node[below=3pt] {23.06}
                   node[above=5pt, text=cyan] {pepflow};
        \draw (8.2,0) node[below=3pt] {23.12}
                   node[below=15pt, text=red] {MMCD};
        \draw (9.5,0) node[below=3pt] {2024.02}
                   node[above=5pt, text=red] {DiffLinker}
                   node[above=20pt, text=cyan] {ForceGen};
        \draw (11,0) node[below=3pt] {24.03}
                   node[below=15pt, text=cyan] {AMP-Diff};
        \draw (12.5,0) node[below=3pt] {24.04}
                   node[above=5pt, text=orange] {RFAA};
        \draw (14,0) node[below=3pt] {24.05}
                   node[below=15pt, text=blue] {Genie2};

        \draw (0.5,-1.5) rectangle ++(0.3,0.3) [fill=red] node[shift={(0.7cm,-0.1cm)}]{EGNN};
        \draw (2.5,-1.5) rectangle ++(0.3,0.3) [fill=orange] node[shift={(1.2cm,-0.1cm)}] {RoseTTAFold};
        \draw (6,-1.5) rectangle ++(0.3,0.3) [fill=blue] node[shift={(0.5cm,-0.1cm)}] {IPA};
        \draw (9,-1.5) rectangle ++(0.3,0.3) [fill=cyan] node[shift={(0.5cm,-0.1cm)}] {ESM};
    \end{tikzpicture}
        \caption{Timeline of major advancements in protein design methods from March 2022 to May 2024. Each event marks the introduction of a significant model or method, categorized by its underlying computational framework. The models are color-coded based on their primary components: Red represents EGNN-based methods, orange corresponds to RoseTTAFold-inspired methods, blue highlights IPA-based methods, and cyan denotes ESM-based methods.}
        \label{fig:timeline_advancements}
\end{figure*}
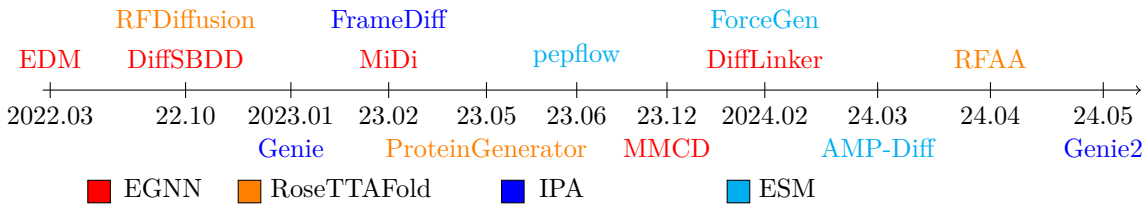

\subsubsection{Model with generalizability}
\label{Diverse}
AlphaFold3 (AF3) \cite{abramsonAccurateStructurePrediction2024} exhibits strong generalizability and versatility, extending beyond protein generation to handle diverse molecular tasks, including ligand and RNA structure prediction.
AlphaFold2 (AF2) \cite{jumper_Highly2021} is a highly accurate protein structure prediction model. Its two important components, Evoformer and IPA, have been widely used in other models. \textbf{AlphaFold3} replaces its Structure Module part with a Diffusion module. The component of the Diffusion module, Diffusion Transformer \cite{peebles_Scalable2023}, shows great generative ability (see Appendix. F. for more
details).\par
Despite the great success of AlphaFold2, AlphaFold3 takes a larger step in this direction. It has many more application scenarios: ligand docking, protein-nucleic acid complexes, covalent modifications, and protein complexes. With AF3, it is possible to handle a more diverse biomolecular space. \par

\section{Diffusion model for peptide generation}

Peptides have aroused great interest due to their potential as therapeutic agents \cite{wang_Therapeutic2022}. Currently, there are several reviews \cite{wan_Deep2022,ge_Editorial2022,goles_Peptidebased2024} that summarize the application of generative models to peptides. Here, we focus on peptide generation by diffusion models.
\par
For the design of peptide sequences, \textbf{ProT-Diff }\cite{wangProTDiffModularizedEfficient2024} combines a pre-trained protein language model (PLM) ProtT5-XL-UniRef50 \cite{elnaggar_ProtTrans2020} with an improved diffusion model to generate \textit{de novo} candidate sequences for antimicrobial peptides (AMPs). \textbf{AMP-Diffusion} \cite{chen_AMPDiffusion2024} uses PLM ESM2 \cite{lin2023evolutionary} for latent diffusion to design AMP sequences with desirable physicochemical properties. This model is versatile and has the potential to be extended to general protein design tasks. 
\textbf{Diff-AMP} \cite{wang_DiffAMP2024} integrates thermodynamic diffusion and attention mechanisms into reinforcement learning to advance research on AMP generation. Sequence-based diffusion models complement structure-based approached by aiding in sequence-to-function or optimizing sequence design for structural goals.\par

For peptide structure design, \textbf{Pepflow} \cite{abdin_PepFlow2023} trains the diffusion model to generate the peptide structure and then uses $E(3)$-equivariant graph neural networks (EGNN) to perform conformational sampling. This model can generate a variety of all-atom conformations for peptides of different lengths, and comparative experiments were performed with AF2 and ESM-fold.
\par

For the co-design of peptides, \textbf{PepGLAD}  \cite{kong_FullAtom2024} proposes geometric latent diffusion model combining with receptor-specific affine transformation to do the full-atom peptide design. \textbf{MMCD} \cite{wangMultiModalContrastiveDiffusion2024} completes the co-generation of structure and sequence for both antimicrobial and anticancer peptides. It also uses EGNN for the structure generation part.\par
All the models for peptide structure generation listed above satisfy the $E(3)$-equivariance property which not only influences the generation of peptides but also provides a guarantee of invariance of the physical properties for binder generation.\par

\section{Small molecule generation}
\begin{figure*}[ht]
\centering
\hspace{2cm}
\includegraphics[bb=-50 10 650 450, width=\textwidth]{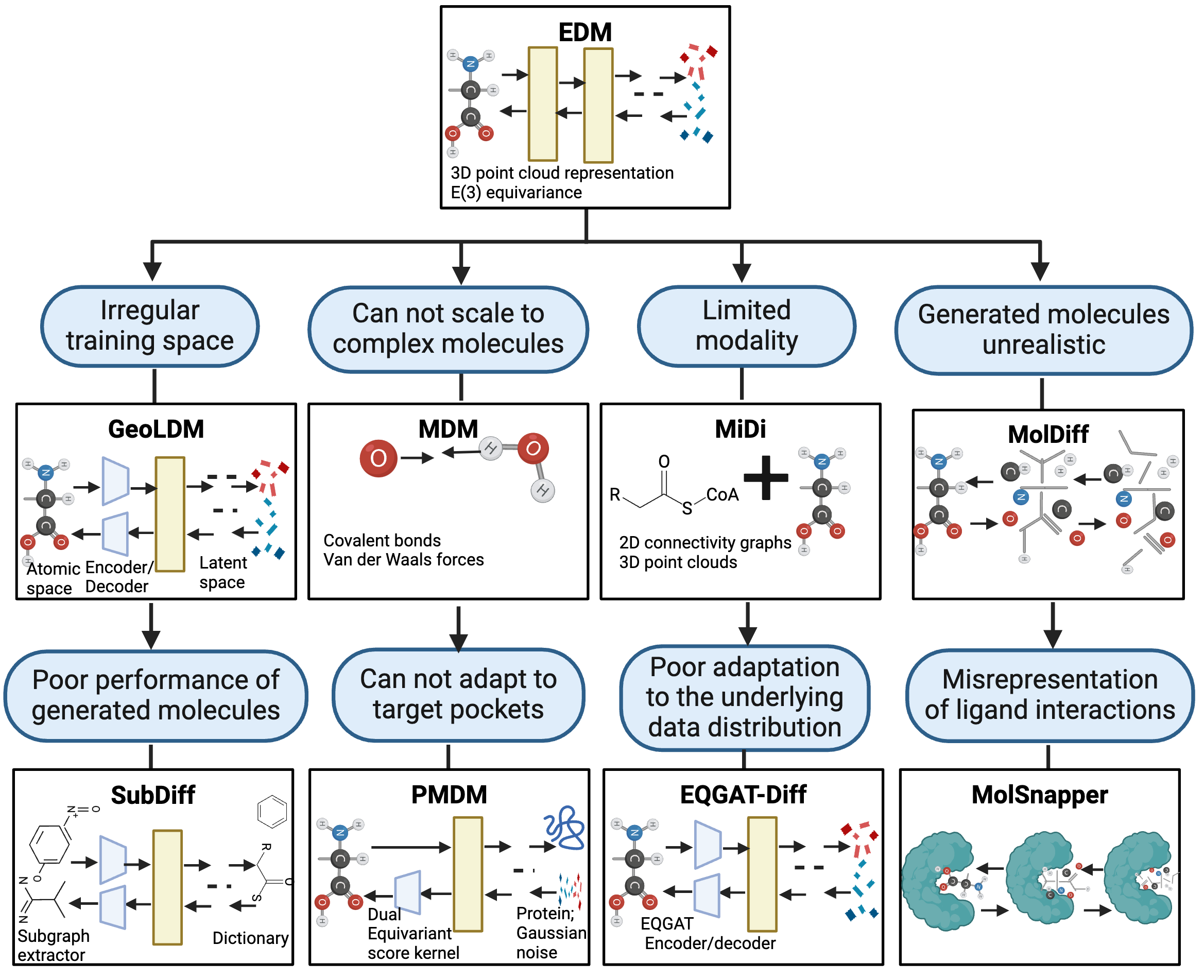}
\caption{Overview of EDM (Equivariant Diffusion Models) and its extensions for molecular generation tasks. The top box represents the foundational EDM model, which uses 3D point cloud representation with E(3) equivariance to handle molecular structures. The figure highlights the key limitations of earlier models (shown in blue boxes). It demonstrates how subsequent models address these challenges through novel methods. Irregular Training Space: GeoLDM uses latent space encoding but performs poorly in generating realistic molecules. SubDiff solves this issue by introducing a subgraph extraction process to improve generation quality. Scalability to Complex Molecules: MDM considers covalent bonds and Van der Waals forces but cannot adapt to target-specific molecular pockets. PMDM incorporates a dual equivariant encoder and Gaussian noise to handle complex protein-ligand interactions. Limited Modality: MiDi combines 2D connectivity graphs and 3D point clouds but struggles with poor adaptation to the data distribution. EQGAT-Diff enhances performance by introducing an EQGAT encoder for better data alignment. Unrealistic Molecules: MolDiff generates molecules with inaccurate ligand interactions. MolSnapper improves molecular realism by accurately representing ligand interactions within target pockets.}
\end{figure*}

Molecules live in physical 3D space, there is a high need to better understand the design space of diffusion models for molecular modeling. The topic of generating molecules using diffusion models is equivalent to the following question: \textit{How to generate attributed graphs using diffusion models?} To answer this question, there are two main challenges:
\begin{itemize}
    \item \textbf{Complex dependency:} Dependency between nodes and edges.
    \item \textbf{Non-unique representations:} Order of the nodes is not fixed.
\end{itemize} \par
For the first challenge, diffusion models need to define the atomic positions $x_{i}\in\mathbb{R}^{3}$ and the atomic types $a_{i}=\{C, N, O,...\}$ and specify independent forward processes for each data type,
\begin{align}
p_{t}(x_{t}|x_{0})&=\mathcal{N}(x_{t}|\alpha_{t}x_{t},\sigma_{t}\mathbf{I}), \\
p_{t}(a_{t}|a_{0})&=\mathcal{N}(a_{t}|\alpha_{t}a_{t},\sigma_{t}\mathbf{I}),
\end{align}
If $G_{t}=(x_{t},a_{t})$, then $p_{t}(G_{t}|G_{0})=\mathcal{N}(x_{t}|\alpha_{t}G_{t},\sigma_{t}\mathbf{I})$, and continuously forward process represented as
\[
d G_{t}=f_{t}(G_{t})dt+g_{t}(G_{t})d\omega_{t},  
\]

The reverse-time diffusion process is represented as:
\begin{equation}
\begin{cases}
dx_{t}=[f_{1,t}(x_{t})-g^{2}_{1,t}\nabla_{x_{t}}\log p_{t}(G_{t})]dt+g_{1,t}d\bar{\omega}_{1},  \\  
da_{t}=[f_{2,t}(x_{t})-g^{2}_{2,t}\nabla_{a_{t}}\log p_{t}(G_{t})]dt+g_{2,t}d\bar{\omega}_{2}.
\end{cases}
\end{equation}
we use $s_{\theta}^{x}(G_{t})$, $s_{\theta}^{a}(G_{t})$ to approximate $\nabla_{x_{t}}\log p_{t}(G_{t})$, $\nabla_{a_{t}}\log p_{t}(G_{t})$ respectively, and train the neural network to jointly approximate the score functions of the constituent processes:
\begin{align}
\mathcal{L}=\mathbb{E}_{x_{t},a_{t}}[&\|s_{\theta}^{a}(G_{t})-\nabla_{a_{t}}\log p_{t}(G_{t}))\| \notag  \\
&+\|s_{\theta}^{x}(G_{t})-\nabla_{x_{t}}\log p_{t}(G_{t})\|]. \notag
\end{align} \par
For the second challenge, diffusion models should capture the system of positional equivariance such as permutation equivariance, $SE(3)$ equivariance and $E(3)$ equivariance. 

\subsection{Permutation equivariant}
\label{permutation}

A model is called equivariant to permutation if its permute input is equivalent to permute output (see more details in Appendix. C.). \par
\textbf{GDSS} \cite{jo_Scorebased2022} is a novel permutation equivariant one-shot diffusion model. It can generate valid molecules by capturing the node-edge relationship. \textbf{CDGS} \cite{huang_Conditional2023} incoporates discrete graph structures into a diffusion model. It is permutation equivariant and implicitly defines the permutation invariant graph log-likelihood function. \par

\textbf{DiGress} \cite{vignac_DiGress2023} is also a permutation equivariant architecture with a permutation invariant loss. The main difference from GDSS is that DiGress defines a diffusion process independent of each node and edge. DiGress achieves better performance than GDSS on QM9 dataset \cite{ramakrishnan_Quantum2014} with simpler architecture. \textbf{JODO} \cite{huang_Learning2023} proposes a diffusion graph transformer to generate 2D graph and 3D geometry molecule generation. Without extra graph structural and positional encoding, JODO-2D is comparable to, or better than, DiGress in most metrics. 

\subsection{Diffusion model on $SE(3)$ group for molecule}
\label{se3_molecule}

We have discussed the important role of the $SE(3)$ equivariant model in protein structure generation before, here we discuss its application in molecule generation. \par
\textbf{GeoDiff} \cite{xuGeoDiffGeometricDiffusion2022} integrates the diffusion model with graph neural networks (GNN) to generate stable conformations, the difference being that the GNN is $SE(3)$-invariant. \textbf{SubGDiff} \cite{zhang_SubGDiff2024} incorporates subgraphs into the diffusion model to improve molecular representation learning. With 500 steps, SubGDiff achieves much better performance than GeoDiff with 5000 steps on 5 out of 8 metrics, which implies that it can accelerate the sampling efficiency. 
\par

Both \textbf{TargetDiff} \cite{guan_3D2023} and \textbf{DiffBP} \cite{lin_DiffBP2024} propose a target-aware molecular diffusion process with a $SE(3)$-equivariant GNN denoiser. The training and sampling procedures in TargetDiff are aligned in non-autoregressive and $SE(3)$ equivariant. DiffBP generates molecules with high protein affinity, appropriate sizes, and favorable drug-like profiles.
\par

\subsection{Models based on EGNNs}
\label{e3_invariant}

We consider the rotation, reflection, and translation group in $\mathbb{R}^{3}$, abbreviated as $E(3)$. 
Since biomolecular structures align with elements in the $E(3)$ group, $E(3)$-equivariant neural networks are effective tools for analyzing molecular structures and properties.
\par

$E(3)$ Equivariant diffusion model (\textbf{EDM}) \cite{hoogeboom2022equivariant} Learns a diffusion model that is equivariant to translation and rotation. It operates on continuous and categorial features to generate molecules in 3D space. \textbf{DiffLinker} \cite{igashovEquivariant3DconditionalDiffusion2024} leverages EDM and develops diffusion models for molecular linker design. \par
\textbf{CGD} \cite{klarner_ContextGuided2024} can consistently generate novel, near-out-of-distribution (near-OOD) molecules with desirable properties. CGD also applies to EDM for material design following the setup of \textbf{GaUDI} \cite{weiss2023guided}, which can discover molecules better than existing ones. \textbf{SILVR} \cite{runcie_SILVR2023} combines ILVR \cite{choi_ILVR2021} and EDM to do fragment merging and linker generation.\par

By building point-structured latent codes with invariant scalars and equivariant tensors, \textbf{GeoLDM} \cite{xu_Geometric2023} can effectively learn latent representations while preserving roto-translational equivariance. It also circumvents the limitations of EDM on irregular training surfaces. \textbf{SubDiff} \cite{yangsubdiff2024} performs subgraph-level encoding in the diffusion process and is used for 3D molecular generation tasks. For unconditional generation tasks, SubDiff is generally better than EDM and GeoLDM. \par

By using a more expressive denoising network, EDM was extended to \textbf{GCDM} \cite{morehead_Geometrycomplete2024}, which margins across conditional and unconditional settings for the QM9 dataset \cite{ramakrishnan_Quantum2014} and the larger GEOM-Drugs dataset \cite{axelrod_GEOM2022}. GCDM is a diffusion model for 3D molecules that can be repurposed for important real-world tasks without retraining or fine-tuning. \textbf{DiffSBDD} \cite{schneuing_Structurebased2023} formulates structure-based drug design (SBDD). \cite{pinheiro_Structurebased2024} follows the noise process in the GCDM. The nodes have both geometric atomic coordinates $x$ as well as nuclear type features $h$. DiffSBDD uses a simple implementation of EGNN to update features $h$ and coordinates $x$.\par 

By limiting the message-passing computations to neighboring nodes, \textbf{MDM} \cite{huang_MDM2022} outperforms EDM in building chemical bonds via atom pair distances. It points out the lack of consideration for interatomic relations in GCDM, and addresses the scalability issue by introducing the Dist-transition Block. \textbf{PMDM} \cite{huangDualDiffusionModel2024} introduces equivariant kernels to MDM to simulate the local chemical boned graph and the global distant graph. \par

\textbf{MiDi} \cite{vignac_MiDi2023} utilizes the adaptive noise schedule and relaxedEGNN (rEGNN) to generate 3D molecules. MiDi outperformed EDM in 2D metrics while obtaining similar 3D metrics for the generated conformers. \textbf{EQGAT-diff} \cite{le_NAVIGATING2024} takes EQGAT \cite{le_Equivariant2022} as the component of the diffusion model to do the \textit{de novo} 3D molecule design. EQGAT-diff employs rotation equivariant vector features that can be interpreted as learnable vector bundles, which the denoising networks of EDM and MiDi are lacking. \par

Taking advantage of the strong relationship between the bond types and bond lengths to guide the generation of atom positions, \textbf{MolDiff} \cite{peng_MolDiff2023} produces high-quality 3D molecular graphs and effectively tackles the atom-bond inconsistency problem with E(3)-equivariant diffusion model. Because it models and diffuses the bonds of molecules, MolDiff surpasses SILVR and EDM in the generation of molecules with better validity. \cite{ziv_MolSnapper2024} extends MolDiff to structure-based drug design and created a model called \textbf{MolSnapper}, which can sample molecules for given pockets. Compared with MolDiff, MolSnapper generates molecules better tailored to fit the given binding site, achieving a high structural and chemical similarity to the original molecules.
\par
A full overview of the developments based on EDM can be seen in Figure 3. The examples above show that the combination of EGNN and diffusion model has been widely used in the generation of proteins, peptides, and small molecules. EGNN is also used alone for protein binding site identification \cite{sestak2024vn}. But EGNN is not always optimal if EGNN and Geometric Vector Perceptron (GVP) are both integrated with \textbf{Keypoint Diffusion}, a diffusion model for \textit{de novo} ligand design: the GVP keypoint model can approach all-atom levels of performance while the EGNN keypoint model failed to exceed the performance $C_{\alpha}$ representation.  \par

\section{Protein-ligand interaction}

\textbf{DiffDock} \cite{corsoDiffDockDiffusionSteps2023} uses an equivariant graph neural network in a diffusion process, and predicts the 3D structure of how a molecule interacts with a protein (shown in Appendix. G). \textbf{DockGen} \cite{corso2024deep} improves upon DiffDock by scaling up the training data and model size, as well as integrating a synthetic data generation strategy based on extracting side chains from real protein structures as ligands. It is faster and better suited for bootstrapping. \par
\textbf{DiffDock-PP} \cite{ketata2023diffdock} learns to translate and rotate unbound protein structures into their bound conformations. \textbf{DiffDock-site} \cite{guo2023diffdock} is a novel paradigm that integrates the precision of the point site for identifying and initializing the docking pocket. It notably outperforms DiffDock in several metrics. Its DiffDock-site-P variant stands out by integrating the pretrained DiffDock for refining ligand attributes. By introducing discrete latent variables to DiffDock, \textbf{DisCo-Diff} \cite{xu_DisCoDiff2024} improves performance on molecular docking and can also synthesise high-resolution images.
\par
\textbf{FABind} \cite{peiFABindFastAccurate2024} takes independent message passing, cross-attention update, and interfacial message passing together, to build a fast and accurate protein-ligand binding model. \textbf{FABind+} \cite{gao_FABind2024} is enhanced by introducing Huber loss in dynamic pocket Radius Prediction and permutation loss in Docking structure prediction.\par

\par
\textbf{NeuralPlexer} \cite{qiaoStatespecificProteinligandComplex2023} incorporates essential biophysical constraints and a multi-scale geometric deep learning system for the diffusion process. For generating the ligand-specific protein-ligand complex structure, a deep equivariant generative model named \textbf{DynamicBind} \cite{lu_DynamicBind2024} is employed. DynamicBind predicts the ligand-specific protein-ligand complex structure with a deep equivariant generative model.\par

All existing deep learning-based methods fail to outperform classical docking tools \cite{buttenschoenPoseBustersAIbasedDocking2024}. Individual data-driven approaches may not provide physically plausible results. We can work towards improving the performance of data-driven deep learning models by introducing physical constraints to diffusion models, such as the model \cite{williams_PhysicsInformed2024}.
\par

\section{Discussion}

Diffusion models have already demonstrated their advantages over previous traditional and machine learning approaches by setting new state-of-the-art results in numerous problems. 
In addition, some basic models have also been frequently used in protein generation recently, such as EGNN, RoseTTAFold, IPA and ESM; these models have derived some new models, which we list in Fig. \ref{fig:timeline_advancements} in the form of timeline.
Here, we highlight several landmark models:

\begin{itemize}
\item The IPA in AlphaFold2 satisfies the property of $SE(3)$ equivariant, but was replaced by the diffusion transformer in \textbf{AlphaFold3}. Therefore, Alphafold3 does not satisfy the properties of an equivariant.

\item The reverse diffusion in \textbf{RFDiffusion} is composed of RoseTTAFold. This model inherits the good properties of RoseTTAFold, making the generated model physically realizable.
\item \textbf{FrameDiff} is the first model to introduce $SE(3)$ manifolds into protein structure generation problems. The properties of the $SE(3)$ group provide a mathematical basis for the expression of structural information. 
\item As a better type of $SE(3)$ equivariant, $E(3)$ equivariant is widely used in the generation of small molecules. The most successful example so far is \textbf{EDM}. 
\item \textbf{DiffDock} is the first model to introduce the use of diffusion models in the molecular docking task, and its performance is very close to traditional methods. Several works proposed different modifications to its framework. 
\end{itemize}
Due to the large size and complexity of protein structures, most current protein models can only satisfy SE(3) equivariance but do not have as good properties as E(3) equivariance. How to establish a diffusion model in the E(3) group to complete protein production is a topic we can study in the future.\par
While progress in the field has demonstrated that diffusion models can accelerate early-stage drug discovery, challenges remain in adapting such workflows to real-world discovery campaigns:
\begin{itemize}
    \item Addressing synthesizability is an ongoing challenge, because many proposed ideas may not have known synthetic routes, and a chemist can only triage a function of proposed ideas.
    \item Despite various widely adopted evaluation metrics, measuring and comparing the performance of diffusion models remains a major challenge given the lack of ground-truth and universal metrics.
    \item Complex dynamics. Cohesive models tend to be static and ignore the fact that proteins and ligands are amphipathic, which is a factor that should be considered when analyzing protein functions. 
    \item Protein structure prediction models typically predict static structures as seen in PDB, not the dynamical behavior of biomolecular systems in solution.
\end{itemize}
\par
\textit{What are potential directions the community could consider exploring further?}
\begin{itemize}
\item  RFDiffusion and ProteinGenerator, which adapt the diffusion model with the traditional model, RoseTTAFold, have done a variety of tasks, such as peptide binder generation, motif-scaffolding, and sequence-structure codesign. We can explore more applications of these two models.
\item There are so few models in the area of diffusion models for peptide design that similar diffusion models for protein design can probably be extended to design peptides. 
\item Traditional models are more analytical and closely match the physical properties of proteins. We can use them for more fruitful tasks such as protein-nucleic acid and protein-ligand interactions.
\item Can ETNN deformations of EGNN \cite{battiloro_Equivariant2024} and NequIP \cite{batzner_Equivariant2022} be applied to the generation of molecules? Can EGNN be used to study peptide structures?
\end{itemize}

\section{Conclusion}
This review comprehensively summarizes the application of the diffusion model for bioengineering. It captures the progression of AI model architectures, highlighting the emergence of $E(3)$ equivariant GNN (EGNN) and diffusion models as game changers in recent work. Diffusion Models are particularly promising generative frameworks.
\section{Acknowledgments}
\label{sec:Acknowledgments}
\renewcommand*{\thesection}{\arabic{section}}
W. Li is supported by a PhD grant from the Region Reunion and European Union (FEDER-FSE 2021/2027)  2023062, 345879. X.F. Cadet is supported by the UKRI CDT in AI for Healthcare http://ai4health.io, (Grant No. P/S023283/1), UK. PEACCEL was supported through a research program partially cofunded by the European Union (UE) and Region Reunion (FEDER). DMO acknowledges ANID for the project "SUBVENCI\'ON A INSTALACI\'ON EN LA ACADEMIA CONVOCATORIA A\~NO 2022", Folio 85220004. DMO gratefully acknowledges support from the Centre for Biotechnology and Bioengineering - CeBiB (PIA project FB0001 and AFB240001, ANID, Chile). MDD acknowledges funding by the Deutsche Forschungsgemeinschaft (DFG, German Research Foundation) - within the Priority Program Molecular Machine Learning SPP2363 (Project Number 497207454). MDD acknowledges EU COST Action CA21160 (ML4NGP).

\bibliographystyle{icml2024.bst}
\bibliography{0_main.bib}

\newpage
\onecolumn
\appendix

\begin{center}
\Large
\textbf{Appendix}
 \\[20pt]
\end{center}
\tableofcontents
\newpage
This is the supplementary material for the review paper "From thermodynamics to protein design: Diffusion models for biomolecule generation towards autonomous protein engineering"
\par
The design of proteins with desirable properties has been a major challenge for biotechnology for decades. Techniques such as directed evolution and rational design have aided protein engineering, but are limited in their explorability. Advances in artificial intelligence have improved on traditional methods and led to semi-rational design and machine learning-assisted directed evolution. Generative approaches such as variational autoencoders and generative adversarial networks have revolutionized biotechnology, but face challenges in the inference and validation of protein structures. \par
Recently, diffusion models have gained interest due to advances in geometric deep learning and computer hardware. These models show high capabilities in generating proteins with stable folding. The review presented here covers diffusion models, geometric deep learning, and matrices of biomolecule generation models. We will present these knowledge backgrounds individually and compare our review with existing ones to help researchers better understand developments in this area.

\addtocontents{toc}{\protect\setcounter{tocdepth}{2}}
\section{Model overview}

\subsection{Glossary}
We describe the terms that are important to our review.
\begin{table}[ht]
\begin{tabular}{|l|l|}
\hline
Terms &
  Description \\ \hline
\begin{tabular}[c]{@{}l@{}}$SE(3)$-\\ equivariance\end{tabular} &
  \begin{tabular}[c]{@{}l@{}}Given an input point cloud $P$, and a random rotation matrix $R$, \\ the network $N$ satisfies $N(PR)=N(P)R$. This kind of network will \\ keep the physical structure.\end{tabular} \\ \hline
\begin{tabular}[c]{@{}l@{}}Atom stability \\ performance\end{tabular} &
  \begin{tabular}[c]{@{}l@{}}Atom stability is calculated as the ratio of atoms exhibiting correct \\ valency, while molecule stability reflects the function of the generated \\ molecules in which each atom maintains stability.\end{tabular} \\ \hline
\begin{tabular}[c]{@{}l@{}}Protein language \\ diffusion model \\ (pLDM)\end{tabular} &
  \begin{tabular}[c]{@{}l@{}}Map the protein sequence to a word-probability latent space.\\ using a pretrained protein language model (pLM)  and train a diffusion \\ model to learn the map between sequence representations.\end{tabular} \\ \hline
\begin{tabular}[c]{@{}l@{}}Physically \\ realizable\end{tabular} &
  \begin{tabular}[c]{@{}l@{}}Designable backbones have optimal secondary structure configurations \\ with favored tertiary structure symmetries such that they are physically \\ realizable with the 20 natural AAs.\end{tabular} \\ \hline
\end{tabular}
\label{Glocary for important terms.}
\end{table}

\subsection{Mindmap for models}
\begin{figure}[ht]
    \centering
    \vspace{5cm}
    \includegraphics[bb=-100 480 3000 1500, width=2\textwidth]{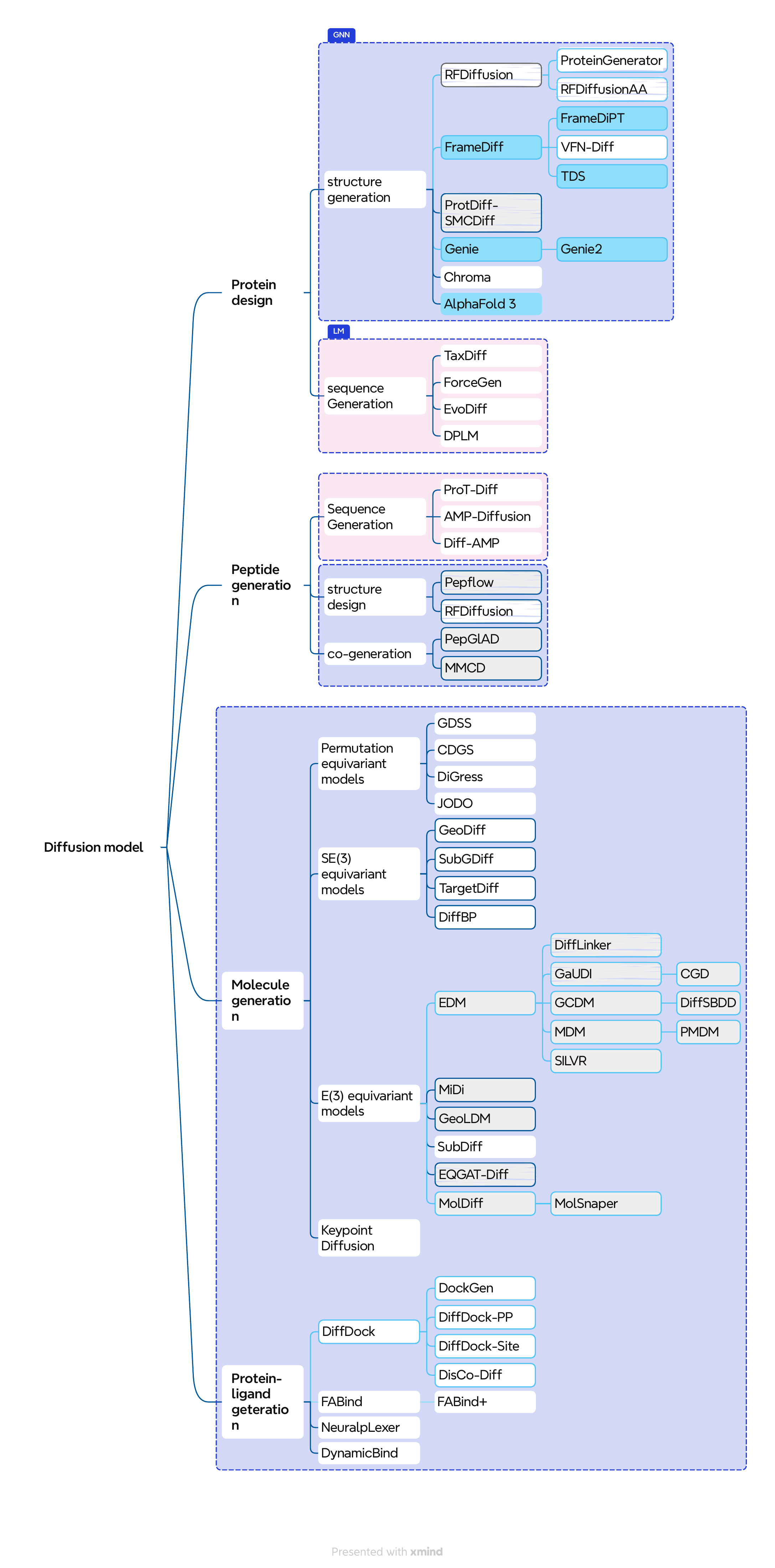}
    \vspace{5cm}
    \caption{Mindmap of the 56 models featured in this review: the models boxed in continuous blue line are the $SE(3)$ equivariant models, the models gray boxes (like Pepflow and PepGLAD) are the $E(3)$ equivariant models, and the blue-shaded ones are models based on Alphafold2. The dark blue branch line indicates the dependence on model classification, and the light blue branch line suggests that the later model is based on the former.}
    \label{mindmap}
\end{figure}

Fig. \ref{mindmap} is a overview of models and the relationships among the models.

\clearpage
\subsection{Model list}
Models mentioned in this review have been implemented as open-source tools. We list their task, input, output, dataset for training, data size, and code link in Table \ref{list}. There are 16 models for protein design, 7 models for peptide generation, 24 models for small molecule generation, and 9 models for protein-ligand interaction, i.e., 56 models in total. This table may help users with their research problems and help developers further improve them.

\begin{longtable}[h]{@{\extracolsep{\fill}}p{1cm}p{1.3cm}p{1.3cm}p{1.3cm}p{1.6cm}p{1.8cm}p{0.7cm}p{0.5cm}cc@{}}

\caption{List of 56 models mentioned in the manuscript with their task, input, output, dataset, data size, code link and reference.} \label{tab:my_label} \\
\hline
\textbf{Task} &\textbf{Paper} & \textbf{Input} & \textbf{output} & \textbf{Dataset} & \textbf{Data Size} & \textbf{Code}  & \textbf{Ref} \\
\hline 
\multirow{18}*{Protein} &\href{https://www.nature.com/articles/s41586-023-06415-8}{RFDiffusion} & structures & structures & PDB  & - & 
\href{https://github.com/RosettaCommons/RFdiffusion}{code} &  \cite{watsonNovoDesignProtein2023} \\
\cline{2-10}
& \href{https://www.biorxiv.org/content/10.1101/2023.10.09.561603v1}{RFAA} & sequence & structures & PDB  & 121,800 & \href{https://github.com/baker-laboratory/rf_diffusion_all_atom}{code} &  \cite{krishna2024generalized} \\
\cline{2-10}
& \href{https://arxiv.org/abs/2302.02277}{FrameDiff}  & structures  & structures &  PDB & 20,312 backbones & \href{https://github.com/jasonkyuyim/se3_diffusion}{code} & \cite{yimSEDiffusionModel2023} \\
\cline{2-10}
&\href{https://www.biorxiv.org/content/10.1101/2023.11.21.568057v1}{FrameDiPT}  & structures  & structures; Full-atom &  RCEB; PDB & 9K clusters & \href{https://github.com/instadeepai/FrameDiPT}{code} & \cite{zhangFrameDiPTSEDiffusion2023} \\
\cline{2-10}
&\href{https://arxiv.org/abs/2306.17775}{TDS}  & structures  & structures &  - & - & \href{https://github.com/blt2114/twisted_diffusion_sampler}{code} & \cite{wu2024practical} \\
\cline{2-10}
&\href{https://arxiv.org/abs/2206.04119}{SMCDiff}  & motif  & scaffolds &  PDB & 4,269 & \href{https://github.com/blt2114/ProtDiff_SMCDiff}{code} & 
\cite{trippe_Diffusion2023} \\
\cline{2-10}
&\href{https://arxiv.org/abs/2310.11802}{VFN-Diff} & structures & structures & PDB & - & \href{https://github.com/aim-uofa/VFN}{code} & \cite{mao_Novo2023} \\
\cline{2-10}
&\href{https://arxiv.org/abs/2301.12485}{Genie} & structures & structures & SCOPe & 195,214 & \href{https://github.com/aqlaboratory/genie}{code} & \cite{lin_Generating2023}\\
\cline{2-10}
&\href{https://arxiv.org/abs/2405.15489}{Genie2} & structures & structures & PDB; AFDB & 588,570 structures & \href{https://github.com/aqlaboratory/genie2}{code} & \cite{lin_Out2024}  \\
\cline{2-10}
&\href{https://www.nature.com/articles/s41586-023-06728-8}{Chroma}  & sequence  & structures   &  PDB, UniProt, PFAM & 28,819 structures & \href{https://github.com/generatebio/chroma}{code}& \cite{ingrahamIlluminatingProteinSpace2023} \\
\cline{2-10}
&\href{https://www.nature.com/articles/s41586-024-07487-w}{AlphaFold3} & sequence; SMILES & structures & PDB 2021 & 41,000,000 structures & \href{https://github.com/google-deepmind/alphafold3}{code} &\cite{abramsonAccurateStructurePrediction2024} \\
\cline{2-10}
&\href{https://www.biorxiv.org/content/10.1101/2023.05.08.539766v1}{PG}  &  sequences  &  structures, sequences  &  - & - & \href{https://github.com/RosettaCommons/protein_generator}{code} & \cite{lisanza2023joint} \\
\cline{2-10}
&\href{https://arxiv.org/abs/2402.17156}{TaxDiff}  & Sequence  & Sequence   &  Alphafold Database; PDB  & - & \href{https://github.com/Linzy19/TaxDiff}{code}& \cite{zongyingTaxDiffTaxonomicGuidedDiffusion2024} \\
\cline{2-10}
&
\href{https://arxiv.org/abs/2310.10605}{Forcegen}  & -  & sequence  &  PDB & 7,026 proteins & - &  \cite{niForceGenEndtoendNovo2024} \\
\cline{2-10}
&\href{https://arxiv.org/abs/2402.18567}{DPLM}  & sequence  & sequence  &  UniRef50 & - & - &  \cite{wang_Diffusion2024} \\
\cline{2-10}
&
\href{https://www.biorxiv.org/content/10.1101/2023.09.11.556673v1}{EvoDiff}  & protein sequences and MSAs  & a new protein sequence  &  OpenFold & - & \href{https://github.com/microsoft/evodiff}{code}& \cite{alamdari2023protein}\\
\hline
\multirow{8}*{Peptide} & \href{https://www.biorxiv.org/content/10.1101/2024.02.22.581480v1.full.pdf}{ProT-Diff}  & sequence & sequence & UniprotKB  & 567,834 peptides  & - &\cite{wangProTDiffModularizedEfficient2024} \\
\cline{2-10}
&\href{https://arxiv.org/pdf/2402.13555}{PepGLAD}  & binding site &  & PDB and literature & - & - & \cite{kong_FullAtom2024}\\
\cline{2-10}
&\href{https://www.biorxiv.org/content/10.1101/2023.06.25.546443v1}{Pepflow}  &  sequences  &  all-atom conformations  & PDB; DBAASP & - & \href{https://gitlab.com/oabdin/pepflow}{code}  & \cite{abdin2023pepflow} \\
\cline{2-10}
&\href{https://www.nature.com/articles/s41586-023-06953-1}{RFDiffusion for peptide} & structures & Designed binder & - & - & \href{https://github.com/RosettaCommons/RFdiffusion}{code} & \cite{vazqueztorresNovoDesignHighaffinity2024} \\
\cline{2-10}
&\href{https://arxiv.org/abs/2312.15665}{MMCD} & sequence; structures  & sequence; structures &  Public databases & 20,129 AMPs; 4,381 ACPs & \href{https://github.com/wyky481l/MMCD}{code} & \cite{wangMultiModalContrastiveDiffusion2024}\\
\cline{2-10}
&\href{https://academic.oup.com/bib/article/25/2/bbae078/7620508}{Diff-AMP} & sequence & sequence & CAMP server & 8,225 AMP sequences & \href{https://github.com/wrab12/diff-amp}{code} & \cite{wang_DiffAMP2024}\\
\cline{2-10}
&\href{https://www.biorxiv.org/content/10.1101/2024.03.03.583201v1.full.pdf}{AMP-Diffusion} & sequence & sequence & dbAMP, AMP Scanner, and DRAMP  & 195,121 peptide sequences & - & \cite{chen2024amp} \\
\hline
\multirow{26}*{Molecule}& \href{https://arxiv.org/abs/2203.17003}{EDM}  &  structures  & structures &  QM9; GEOM-Drugs & 100K & - & \cite{hoogeboom2022equivariant} \\
\cline{2-10}
&
\href{https://arxiv.org/abs/2209.05710}{MDM}  &  geometries  & geometries &  QM9; GEOM & 290K & - & \cite{huang_MDM2022}\\
\cline{2-10}
&
\href{https://arxiv.org/abs/2302.04313}{GCDM}  &  3D graph  & 3D graph &  QM9; GEOM-Drugs & 100K & \href{https://github.com/BioinfoMachineLearning/Bio-Diffusion}{code} & \cite{morehead_Geometrycomplete2024} \\
\cline{2-10}
&
\href{https://arxiv.org/abs/2210.13695}{DiffSBDD}  &  pockets  & ligands &  CrossDocked; Binding MOAD & - & \href{https://github.com/arneschneuing/DiffSBDD}{code} & \cite{schneuing_Structurebased2023}\\
\cline{2-10}
&
\href{https://arxiv.org/abs/2305.01140}{GeoLDM}  &  geometries  & structures &  QM9; GEOM-Drugs & - & \href{https://github.com/MinkaiXu/GeoLDM}{code} & \cite{xu_Geometric2023}\\
\cline{2-10}
&
\href{https://arxiv.org/abs/2302.09048}{MiDi}  & graph structures  & graph &  QM9; GEOM-Drugs & - & \href{https://github.com/cvignac/MiDi}{code} & 
\cite{vignac_MiDi2023}\\
\cline{2-10}
&
\href{https://www.nature.com/articles/s42256-024-00815-9}{DiffLinker}  & structures & Molecule structures &  ZINC, CASF, GEOM & 185,678 examples & \href{https://github.com/igashov/DiffLinker}{code} & \cite{igashovEquivariant3DconditionalDiffusion2024} \\
\cline{2-10}
&
\href{https://www.nature.com/articles/s41467-024-46569-1}{PMDM} & Molecule, protein pocket  & molecule structures  &  CrossDocked  & 22.5 million docked protein-ligand pairs & \href{https://github.com/Layne-Huang/PMDM/tree/main}{code}& \cite{huangDualDiffusionModel2024} \\
\cline{2-10}
&
\href{https://arxiv.org/abs/2309.17296}{EQGAT-Diff} & structures  & molecule structures  &  QM9; GEOM-Drugs; CrossDocked; PubChem3D  & - & \href{https://github.com/tuanle618/eqgat-diff}{code}& \cite{le2023navigating} \\
\cline{2-10}
&
\href{https://arxiv.org/abs/2211.11214}{DiffBP} &  binding site  & molecule structures  &  CrossDocked  & 10,000 protein-ligand paired samples & - & \cite{lin_DiffBP2024}\\
\cline{2-10}
&
\href{https://arxiv.org/abs/2311.13466}{Keypoint Diffusion} & molecule structures & ligands & BindingMOAD & 40,000 & \href{https://github.com/dunni3/keypoint-diffusion}{code} & \cite{dunn2023accelerating} \\
\cline{2-10}
&
\href{https://arxiv.org/abs/2203.02923}{Geodiff} & molecular graphs & molecular conformations & QM9; GEOM-Drugs & 200,000 conformations & \href{https://github.com/MinkaiXu/GeoDiff}{code} & \cite{xuGeoDiffGeometricDiffusion2022}  \\
\cline{2-10}
&
\href{https://arxiv.org/abs/2303.03543}{TargetDiff} & binding site & binding molecules & CrossDocked2020 & 100,000 complexes & \href{https://github.com/guanjq/targetdiff}{code} & \cite{guan_3D2023} \\
\cline{2-10}
&
\href{https://arxiv.org/abs/2305.07508}{MolDiff} & molecular structures & molecular structures & QM9; GEOM-Drugs &231,523 molecules & \href{https://github.com/pengxingang/MolDiff}{code} & \cite{peng_MolDiff2023} \\
\cline{2-10}
&
\href{https://www.biorxiv.org/content/10.1101/2024.03.28.586278v1}{MolSnapper} & Protein-ligand complex & molecules & CrossDocked; Binding MOAD & - & \href{https://github.com/oxpig/MolSnapper}{code} & \cite{ziv2024molsnapper} \\
\cline{2-10}
&
\href{https://arxiv.org/abs/2407.11942}{CGD} & molecule graph & molecule graph & Zink & 250 000 small molecules & 
\href{https://github.com/leojklarner/context-guided-diffusion}{code} & \cite{klarner_ContextGuided2024} \\
\cline{2-10}
&
\href{https://arxiv.org/abs/2202.02514}{GDSS} & graph structures & structures & QM9 and ZINC250k & 10,000 molecules & \href{https://github.com/harryjo97/GDSS}{code} & \cite{jo_Scorebased2022} \\
\cline{2-10}
& \href{https://arxiv.org/abs/2301.00427}{CDGS} & graph & graph & ZINC250k; QM9 & 383,340 molecules & \href{https://github.com/GRAPH-0/CDGS}{code} & \cite{huang_Conditional2023} \\
\cline{2-10}
& \href{https://arxiv.org/abs/2209.14734}{DiGress} & graph & atomic coordinates & MOSES and GuacaMol & - & \href{https://github.com/cvignac/DiGress}{code} & \cite{vignac_DiGress2023}\\
\cline{2-10}
& \href{https://arxiv.org/abs/2305.12347}{JODO} & graph & graph & QM9; GEOM-Drugs; ZINC250k; MOSES & 2,621,542 molecules & \href{https://github.com/GRAPH-0/JODO}{code} & \cite{huang_Learning2023} \\
\cline{2-10}
& \href{https://arxiv.org/abs/2405.05665}{SubGDiff} & molecular graph & graph & PCQM4Mv2 & 3.4 million molecules & \href{https://github.com/IDEA-XL/SubgDiff}{code} & \cite{zhang_SubGDiff2024} \\
\cline{2-10}
& \href{https://www.nature.com/articles/s43588-023-00532-0}{GaUDI} & molecular graph & graph & cc-PBH; PAS & 509,000 molucules & \href{https://gitlab.com/porannegroup/gaudi}{code} & \cite{weiss2023guided} \\
\cline{2-10}
& \href{https://arxiv.org/abs/2304.10905}{SILVR} & multiple superimposed fragments & graph & COVID Moonshot dataset & - & \href{https://github.com/nichrun/e3_diffusion_for_molecules}{code} & \cite{runcie_SILVR2023} \\
\cline{2-10}
& \href{https://openreview.net/forum?id=z2avrOUajn}{SubDiff} & subgraph & generative graph & GEOM-Drug; QM9 & - & - & \cite{yangsubdiff2024}\\
\hline
\multirow{9}*{\makecell{Protein-\\ligand}} & \href{https://arxiv.org/pdf/2210.01776}{DiffDock}  & Ligand and protein structures  & ligand pose distributions&  PDBBind & - & \href{https://github.com/gcorso/DiffDock}{code} & \cite{corsoDiffDockDiffusionSteps2023} \\
\cline{2-10}
&
\href{https://arxiv.org/abs/2304.03889}{DiffDock-PP}  & protein structures  & complex structures &  DIPS  & 42,826 & \href{https://github.com/ketatam/DiffDock-PP}{code} &  \cite{ketata2023diffdock} \\
\cline{2-10}
&
 \href{https://arxiv.org/abs/2209.15171}{Neural-PLexer}  & Protein Sequences; ligand Molecular Graphs & Complex Structures   &  PL2019-74k, PDBBind2020; PocketMiner ;GPCRdb & 74,477 samples & \href{https://github.com/zrqiao/NeuralPLexer}{code}& \cite{qiaoStatespecificProteinligandComplex2023} \\
\cline{2-10}
&
\href{https://openreview.net/pdf?id=AlPg6if5PU}{DiffDock-Site}  & protein structures  & ligand structures &  PDBBind & 17,000 complexes & - &  \cite{guo2023diffdock} \\
\cline{2-10}
&
\href{https://arxiv.org/abs/2407.03300}{DiSco-Diff}  & molecular structures  & molecular structures &  PDBBind & - & \href{https://research.nvidia.com/labs/lpr/disco-diff}{code} & \cite{xu_DisCoDiff2024} \\
\cline{2-10}
&
\href{https://arxiv.org/abs/2402.18396}{DockGen}  & protein and ligand structures  &  ligand &  PDBBind; Binding MOAD & - & \href{https://github.com/gcorso/DiffDock}{code} & \cite{corso2023discovery} \\
\cline{2-10}
&
\href{https://arxiv.org/abs/2310.06763}{FABind}  & protein-ligand complex &  binding pose of the ligand &  PDBBind & 17,299 complexes & \href{https://github.com/QizhiPei/FABind}{code} & \cite{peiFABindFastAccurate2024} \\
\cline{2-10}
&
\href{https://arxiv.org/abs/2403.20261}{FABind+}  & protein-ligand graph &  binding pose of the ligand &  PDBBind & 17,644 samples & - & \cite{gao_FABind2024} \\
\cline{2-10}
&
\href{https://www.nature.com/articles/s41467-024-45461-2}{Dynamic- Bind}  & protein structures &  ligand and protein residues &  PDBBind &19,443 crystal structures & \href{https://github.com/luwei0917/DynamicBind}{code} & \cite{lu_DynamicBind2024} \\
        \hline
\label{list}
\end{longtable}

\section{Extension of Diffusion Models}
We only introduced the representation of the diffusion model in the main text. In this part, we supplement its derivation, application, and improvement.
\subsection{Loss function of DDPM}
\begin{definition}
    (KL divergence) \cite{li_Generalization2024a} Given two distributions $p$ and $q$, the KL divergence from $q$ to $p$ is defined as $D_{kL}=\int_{\mathbb{R}^{d}}p(x)\frac{p(x)}{q(x)}dx$.
\end{definition}
The VAE \cite{wei2020recent} loss is a bound on the true log-likelihood (also called the variational lower bound):
\[
-L_{VAE}=\log p_{\theta}(x)-D_{kL}(q_{\phi}(z|x)\|p_{\theta}(z|x))\leq \log p_{\theta}(x).
\]
Apply the same trick to diffusion model:
\[
-\log p_{\theta}(x_{0})\leq \mathbb{E}_{q(x_{0:T})}[-\log \frac{p_{\theta}(x_{0:T})}{q(x_{1:T}|x_{0})}]=L_{VLB}.
\]
Expanding out,
\begin{align}
L_{VLB}&=L_{reconstruct}+L_{prior}+L_{denoise},  \notag\\
\text{where} \qquad L_{prior}&=D_{kL}(q(x_{T}|x_{0})\|p_{\theta}(x_{T})),  \notag\\
L_{denoise}&=\sum_{t=2}^{T}D_{kL}(q(x_{t}|x_{t+1},x_{0})\|p_{\theta}(x_{t}|x_{t+1})), \notag\\
L_{reconstruct}&=-\log p_{\theta}(x_{0}|x_{1}). \notag
\end{align}

\subsection{Conditional Diffusion Models}

Denote the conditional information as $y$, the goal of conditional diffusion models is to generate samples from the conditional data distribution $p(\cdot|y)$. The conditional forward process can be written as:
\begin{equation}
dX_{t}^{y}=-\frac{1}{2}x_{t}^{y}dt+d\omega_{t} \qquad \text{with } x_{0}^{y}\sim p_{0}(\cdot |y) \text{ and }t\in[0,T]. 
\label{conditional}
\end{equation}
Similarly, for sample generation, the backward process reverses the time in \eqref{conditional}:
\[
dx_{t}^{y}=[\frac{1}{2}x_{t}^{y}+\nabla \log p_{T-t}(x_{t}^{y}|y)]dt+d\bar{\omega}_{t}, \text{  for }t\in[0,T),
\]
here $\nabla \log p_{T-t}(x_{t}^{y}|y)$ is the so-called conditional score function.

\section{Compare with other models}
Diffusion models have surpassed the previous dominant generative adversarial networks (GANs) in the challenging task of image synthesis. In particular, the computational resource requirements are much higher than those of GAN and VAE. Although diffusion models tend to be slow at generating samples, their generation of high-fidelity and high diversity samples has made them a popular choice for recent protein engineering applications.

\subsection{Pros, cons, and developments of diffusion models}
Diffusion models have some advantages over other models:
\begin{itemize}
    \item They give amazing results for image, audio and text synthesis, while being relatively simple to implement.
    \item They are related to stochastic differential equations (SDEs), and thus their theoretical properties are particularly intriguing.

\end{itemize}
But diffusion models also have some technical shortcomings:
\begin{itemize}
    \item They cannot learn the representation of biomolecules, and need the help of graph neural networks or large language models to complete the representation of structures or sequences.
    \item No dimensional changes. The dimensionality of input is kept across the whole model.
\end{itemize}

\par
Developments for enhancing diffusion models \cite{cao2024survey}: (1) speed up the standard Ordinary differential equation (ODE) or SDE simulation; (2) improve Brownian motion in pixel space; (3) enhance the diffusion ODE likelihood; (4) bridging distribution techniques that utilize diffusion model concepts to connect two distinct distributions.

\section{Permutation equivariance}
Permutation equivariance is an important concept in geometric graph neural networks. It is also used to generate molecules in diffusion models. We provide a detailed knowledge of it here.
\subsection{Permutation Equivariant}
\begin{figure}[ht]
    \centering
    \includegraphics[bb=-200 0 450 180, width=\textwidth]{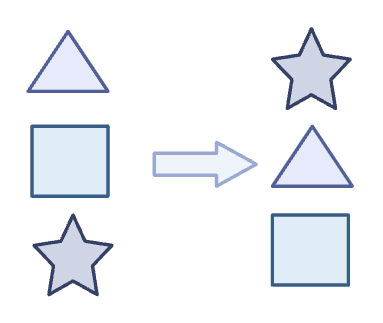}
    \caption{Permutation is reordering the elements in a group. }
    \label{permutation_element}
\end{figure}
 A group is a set $G$ with a binary operation $G\times G\rightarrow G$ denoted $gh$ satisfying the following properties:
    \begin{itemize}
    \item Closure: $gh\in G$ for all $g,h\in G$;
    \item Associativity: $(gh)l=g(hl)$ for all $g,h,l\in G$;
    \item Identity: there exists a unique $e\in G$ satisfying $eg=ge=g$;
    \item Inverse: for each $g\in G$, there is as unique inverse $g^{-1}\in G$, such that $gg^{-1}=g^{-1}g=e$.
    \end{itemize}
Permutation equivariance in a group means that permutation of the elements of a set (see Fig. \ref{permutation_element}) preserves set membership. \par
A graph $G$ with $N$ nodes is defined by its node features $x\in \mathbb{R}^{N\times F}$ and the weighted adjacency matrix $A\in \mathbb{R}^{N\times N}$ as $(x,A)\in \mathbb{R}^{N\times F}\times\mathbb{R}^{N\times N}:= \mathcal{G}$, where $F$ is the dimension of the node features. Permutation equivariance for a graph implies that any permutation of the columns of $x$ and $A$ is equiprobable.

\begin{theorem}
    (Permutation equivariance of graph neural network) Consider consistent permutations of the shift operator $\hat{s}=P^{T}sP$ and input signal $\hat{x}=P^{T}x$. Then 
    \[
    \phi(\hat{x};\hat{s},H)=P^{T}\phi(x;s,\mathcal{H}).
    \]
\end{theorem}

\subsection{Relationship between Permutation Invariance and Permutation Equivariance}

\begin{figure}[ht]
    \centering
    \includegraphics[bb=0 0 600 450, width=8cm]{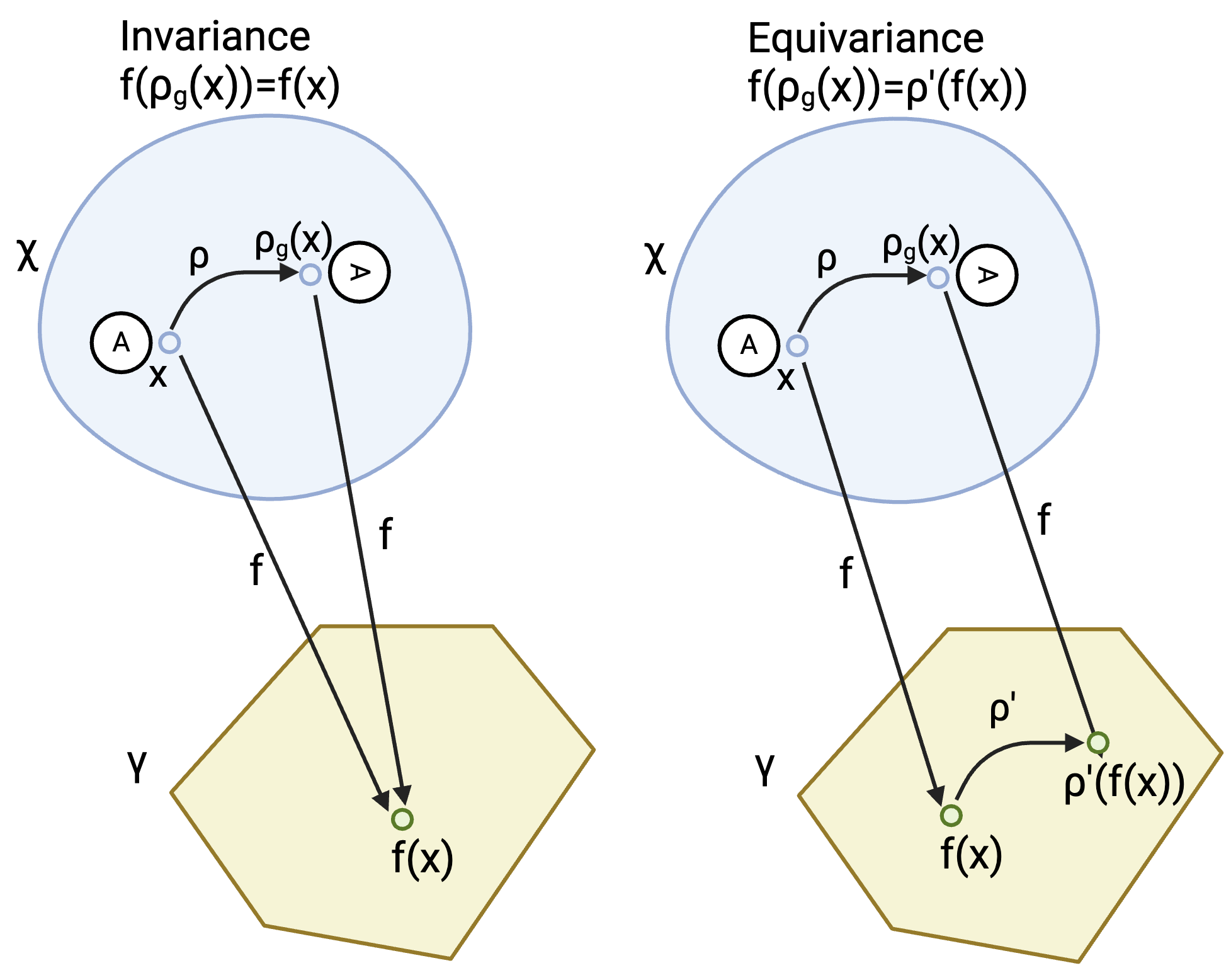}
    \caption{Invariance and equivariance. }
    \label{invariance}
\end{figure}

Equivariance and invariance are two similar concepts, but lead to entirely different implications (see Fig. \ref{invariance}). 

For invariance, we expect the output to remain completely unchanged regardless of changes in input: $f(\rho_{g}(x))=f(x)$.\par

\begin{definition}
    (Permutation Operation on Matrix) Let $[N]\triangleq\{1,...,N\}$. Denote the set of permutations $\pi:[N]\Rightarrow[N]$ as $\Pi_{N}$. The node permutation operation on a matrix $A\in\mathbb{R}^{N\times N}$ is defined by $A_{i,j}^{[\pi]}=A_{\pi(i),\pi(j)}$.
\end{definition}
\begin{definition}
(Permutation Invariant Function) A function $f$ with $\mathbb{R}^{N\times N}$ as its domain is permutation invariant if $\forall A \in \mathbb{R}^{N\times N}, \forall \pi \in \Pi_N, \quad f(A^{[\pi]}) = f(A)$.
\end{definition}

\begin{lemma}
	(Permutation Invariance of Frobenius Inner Product) For any $A, B \in \mathbb{R}^{N\times N}$, the Frobenius inner product of $A,B$ is $\langle A, \mathbf{B}\rangle_{\mathrm{F}}=\sum_{i, j} {A_{i j}} B_{i j}=\operatorname{tr}({A^{T}} \mathbf{B})$. Frobenius inner product operation is permutation invariant, i.e., $\forall \pi \in \Pi_N, \quad \langle A^{[\pi]}, \mathbf{B}^{[\pi]}\rangle_{\mathrm{F}} = \langle A, \mathbf{B}\rangle_{\mathrm{F}}$. %
\end{lemma}

\begin{theorem} 
(Relationship between permutation equivariance and invariance) \cite{niu_Permutation2020} If $\mathbf{s}: \mathbb{R}^{N\times N} \rightarrow \mathbb{R}^{N\times N}$ is a permutation equivariant function, then the scalar function $f_{\mathbf{s}}= \int_{\gamma[\mathbf{0}, A]}\langle\mathbf{s}(\mathbf{X}), \operatorname{d}\mathbf{X}\rangle_{\mathrm{F}} + C$ is permutation invariant, where $\langle A, \mathbf{B}\rangle_{\mathrm{F}}=\operatorname{tr}({A^T} \mathbf{B})$ is the Frobenius inner product, $\gamma[\mathbf{0}, A]$ is any curve from $\mathbf{0} = \{0\}_{N\times N}$ to $A$, and $C \in \mathbb{R}$ is a constant.
\end{theorem}

\section{$SE(3)$ Group}
\label{SE3}

$SE(3)$ equivariance has been repeatedly mentioned in the review and plays an important role in the structure generation of proteins, peptides, small molecules, and protein ligands. Here we give its definition, properties, and relationship with the protein framework. 

\subsection{Definition of $SO(3)$ and $SE(3)$}
In mathematics, a rigid can be abstracted into 3-dimensional geometric coordinates. Its position can be represented by a matrix of $n$ coordinates, and operations such as rotating and translating the object are equivalent to multiplying the coordinates by a 3-dimensional matrix. i.e.,
\[
\begin{pmatrix}
x_1 & y_2 & z_{1} \\
\vdots & \vdots &\vdots \\
x_{n} & y_{n} & z_{n} \\
\end{pmatrix}
\times
\underbrace{
\begin{pmatrix}
a_1 & b_1 & c_1\\
a_2 & b_2 & c_2\\
a_3 & b_3 & c_3\\
\end{pmatrix}}_{\text{denoted as } A}
=
\begin{pmatrix}
x'_1 & y'_2 & z'_{1} \\
\vdots & \vdots &\vdots \\
x'_{n} & y'_{n} & z'_{n} \\
\end{pmatrix}
\]
\begin{definition} \label{representation}
    (Representation) An n-dimensional real representation of a group $G$ is a map $\rho:G\rightarrow \mathbb{R}^{n\times n}$, assigning to each $g\in G$ an invertable matrix $\rho(g)$, and satisfying
\[
\rho(e)=1, \quad \rho(gh)=\rho(g)\rho(h), \qquad \forall g,h\in G.
\]
\end{definition}

According to Definition \ref{representation}, a group can be uniquely represented by a matrix. Here, we introduce several general groups. A general linear group $GL(n)$ is the group of invertible $n\times n$ matrices. A matrix $A\in GL(n)$ is orthogonal if $Av\cdot Aw=v\cdot w$ for all vectors $v$ and $w$. If $A$ is an orthogonal matrix, it is equivalent to a rotation transformation of the object. We mark this type of transformation as O(3). If $A$ is an orthogonal matrix and its determinant is 1, it is called a special orthogonal matrix. The corresponding transformation is the combination of rotation and reflection transformation, denoted as SO(3). \par

The orthogonal matrices form a subgroup $O(n)$ of $GL(n)$. The orthogonal matrices with determinant 1 form a subgroup $SO(n)\subset O(n)\subset GL(n)$ called the special orthogonal group. 

The special Orthogonal group in 3 dimensions consists of the 3D rotation matrices:
\[
SO(3)=\{\gamma\in\mathbb{R}^{3x3}: \gamma^{T}\gamma=\gamma\gamma^{T}=\mathbf{I}, det \gamma=1\}.
\]
Inner product and distance on $SO(3)$:
\[
\langle\gamma_{1},\gamma_{2}\rangle_{SO(3)}=\frac{1}{2}tr(\gamma_{1}^{T},\gamma_{2}) \qquad
d_{SO(3)}(\gamma_{1},\gamma_{2})=\|\log (\gamma_{1}^{T}\gamma_{2})\|_{F}.
\]
The special Euclidean group, $SE(3)$ is used to represent rigid body transformations in 3D:
\[
SE(3)=\Biggr\{A|A=
\begin{bmatrix}
    \gamma & s \\
    0 &1
\end{bmatrix}: \gamma\in SO(3), s\in (\mathbb{R}^{3},+)\Biggr\}.
\]
Inner product and distance on $SE(3)$:
\[
\langle A_{1}, A_{2} \rangle_{SE(3)}=\langle \gamma_{1},\gamma_{2} \rangle_{SO(3)}+\langle s_{1},s_{2}\rangle_{\mathbb{R}^{3}}
\]
\[
d_{SE(3)}(A_{1}, A_{2})=\sqrt{d_{SO(3)}(\gamma_{1},\gamma_{2})^{2}+d_{\mathbb{R}^{3}}(s_{1},s_{2})^{2}}.
\]
$SE(3)$ satisfies the following four axioms:
\begin{itemize}
    \item The set is closed under the binary operation. $A, B\in SE(3) \Rightarrow AB\in SE(3)$.
    \item The binary operation is associative. $A,B,C\in SE(3)\Rightarrow (AB)C=A(BC)$.
    \item $\exists I\in SE(3)$, s.t., $\forall A\in SE(3)$, $AI=A$.
    \item $\forall A\in SE(3)$, $\exists A^{-1}\in SE(3)$, $AA^{-1}=I$.
\end{itemize}
\begin{definition}
    (Manifold) A manifold of dimension $n$ is a set $M$ locally homeomorphic to $\mathbb{R}^{n}$.
\end{definition}
\begin{definition}
    (Lie Group) A Lie group is a topological group that is also a differentiable manifold and such that the composition and inverse operations $G\times G\rightarrow G$ and $G\rightarrow G$ are infinitely differentiable functions.
\end{definition}

$SE(3)$ is a continuous group, and the open set of elements of $SE(3)$ has 1-1 map onto an open set of $\mathbb{R}^{6}$. In other words, $SE(3)$ is a differentiable manifold, i.e., a Lie group.  \par

\subsection{$SE(3)$ for the representation of protein frame}

\begin{figure}[ht]
    \centering
    \includegraphics[bb=0 0 250 250, width=4cm]{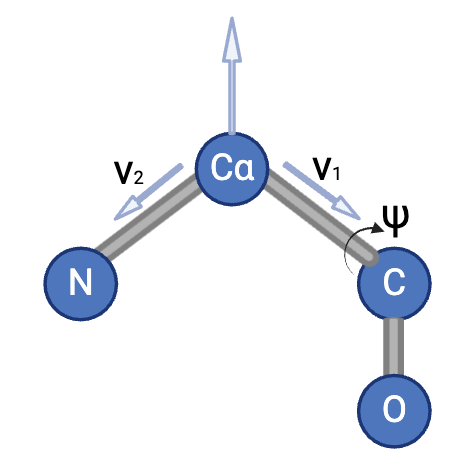}
    \caption{Visualization of a protein frame.}
\end{figure}
For a protein frame, the atomic coordinates can be defined as:
\[
[N_{n}, C_{n},(C_{\alpha})_{n}]=[T_{n}]\cdot [N^{*},C^{*},C_{\alpha}^{*}],
\]
$n$ is the index of residue, $r_{n}=$Gram-Schmidt$(v_{1},v_{2})$, $x_{n}=C_{\alpha}\in\mathbb{R}^{3}$, $T_{n}=(r_{n},x_{n})$ is a member of the special Euclidean group $SE(3)$.\par
Advantage to use $SE(3)$ equivariant graph neural networks for protein generation:
\begin{itemize}
    \item Leverage symmetries can improve sample efficiency, reduce complexity, and enhance generalizability.
    \item Respect geometrical or physical constraints.
    \item Interpretability. A model constrained by the symmetry group(s) of the underlying problem may be more interpretable than a general one, not only because it is likely to have fewer parameters but also because these parameters will represent physically meaningful observables. 
\end{itemize}

\section{EGNNs}
In this section, we present Equivariant Graph Neural Networks (EGNNs) and demonstrate their definition and relationship with $SE(3)$ equivariant neural network.
\subsection{Relationship between $SE(3)$ and $E(3)$}
\begin{figure}[ht]
        \includegraphics[bb=-50 0 800 420, width=\textwidth]{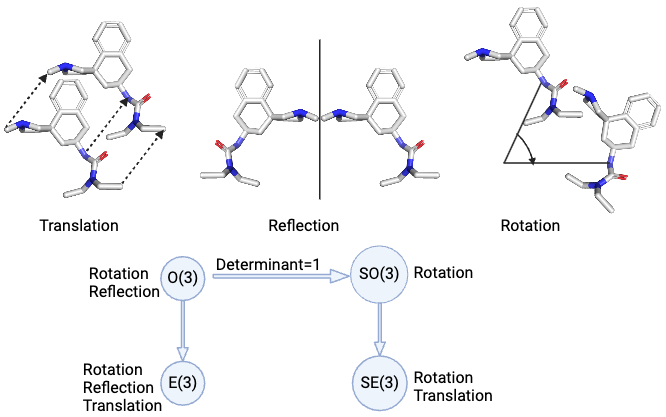}
        \caption{Information about $E(3)$ group. Top subplot: Illustration of translation, reflection and rotation. Bottom subplot: Relationship among several general groups.}
        \label{translation}
    \label{workflow}
\end{figure}

$E(3)$ is the notation for the Euclidean group that denotes the set of isometric transformations in Euclidean space, and the transformations in $E(3)$ include translation, rotation, and reflection. The relationship between $SE(3)$ and $E(3$) (see Fig. \ref{translation}): $SE(3)$ is a subgroup of $E(3)$ that includes only translation and rotation, while $E(3)$ includes the motions of $SE(3)$ as well as reflective motions. In other words, $SE(3)$ is a subset of $E(3)$ that remains oriented.\par

$E(3)$ equivariant neural networks are computationally more efficient and have been shown to perform equal to, or better than, $SE(3)$ equivariant networks for the modelling of quantum chemical properties and dynamic systems \cite{bogatskiy2022symmetry}.
\subsection{Concept of EGNN}

A point cloud is a finite set of $3D$ coordinates where every point has a corresponding feature vector. A function $f$ is $E(3)$-equivariant if for any point cloud $x$, orthogonal matrix $R\in \mathbb{R}^{3\times 3}$ and translation vector $t\in \mathbb{R}^{3}$ we have: $f(Rx+t)=Rf(x)+t$. A conditional distribution $p(x|y$) is $E(3)$-equivariant if for any point clouds $x,y$, $p(Rx+t|Ry+t)=p(x|y)$. A function $f$ and a distribution $p$ are $O(3)$-equivariant if $f(Rx)=Rf(x)$ and $p(Rx|Ry)=p(x|y)$. 
\begin{definition}
    An equivariant neural network is a neural network in which each layer is a direct sum of permutation representation representations, and all linear maps are G-equivariant.
\end{definition}

Equivariant graph neural networks (EGNNs) according to the definition of \cite{satorras_Equivariant2022} equivariant to the Euclidean group $E(3)$ of rigid motions (rotations, translations, and reflections) in addition to the standard permutation equivariance.
EGNNs have attracted attention in the natural and medical sciences because they represent a new tool for analyzing molecules and their properties.
Reasons for the use of EGNNs:
\begin{enumerate}
\item[$\bullet$] Rotations and translations in 3D space act on the entire input set of particles and lead to the same translations on their entire trajectory.
\item[$\bullet$] No matter which model we use, we will have to generalize over the equivariant of this task to achieve good performance.
\item[$\bullet$] EGNN is more data efficient than GNN since it does not require generalization over rotations and translations of the data.
\end{enumerate}
EGNNs are computationally efficient and easy to implement, but are limited in use cases and sometimes hard to abstract real-world scenarios into a coordinate system.
\par

\section{AlphaFold3}
\begin{figure}[ht]
    \centering
    \vspace{0.5cm}
    \begin{minipage}{0.6\textwidth}
        \includegraphics[bb=0 0 800 600, width=\textwidth]{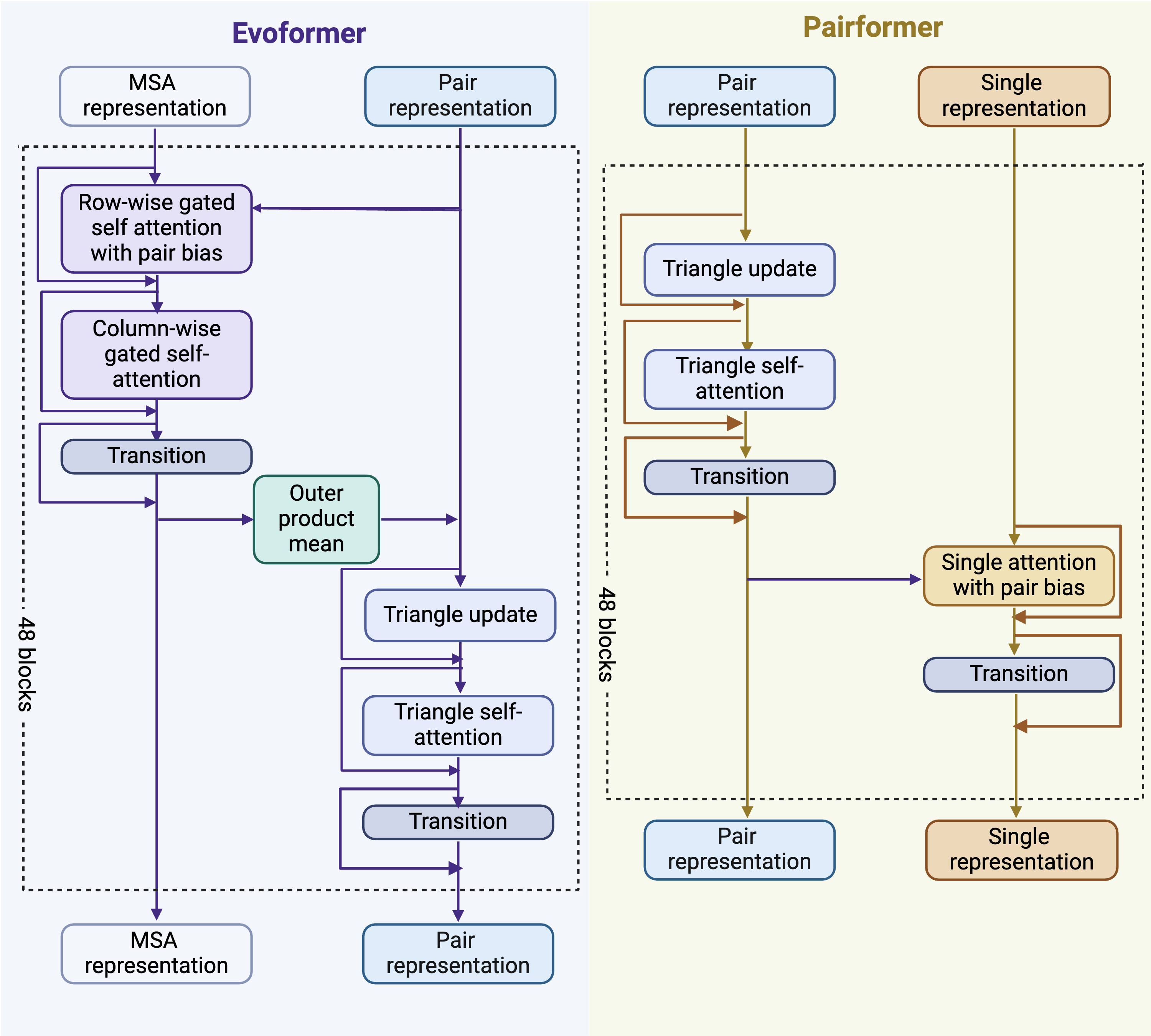}
    \end{minipage}
    \vspace{1.5cm}
       \begin{minipage}{0.6\textwidth}
        \includegraphics[bb=0 100 800 700, width=\textwidth]{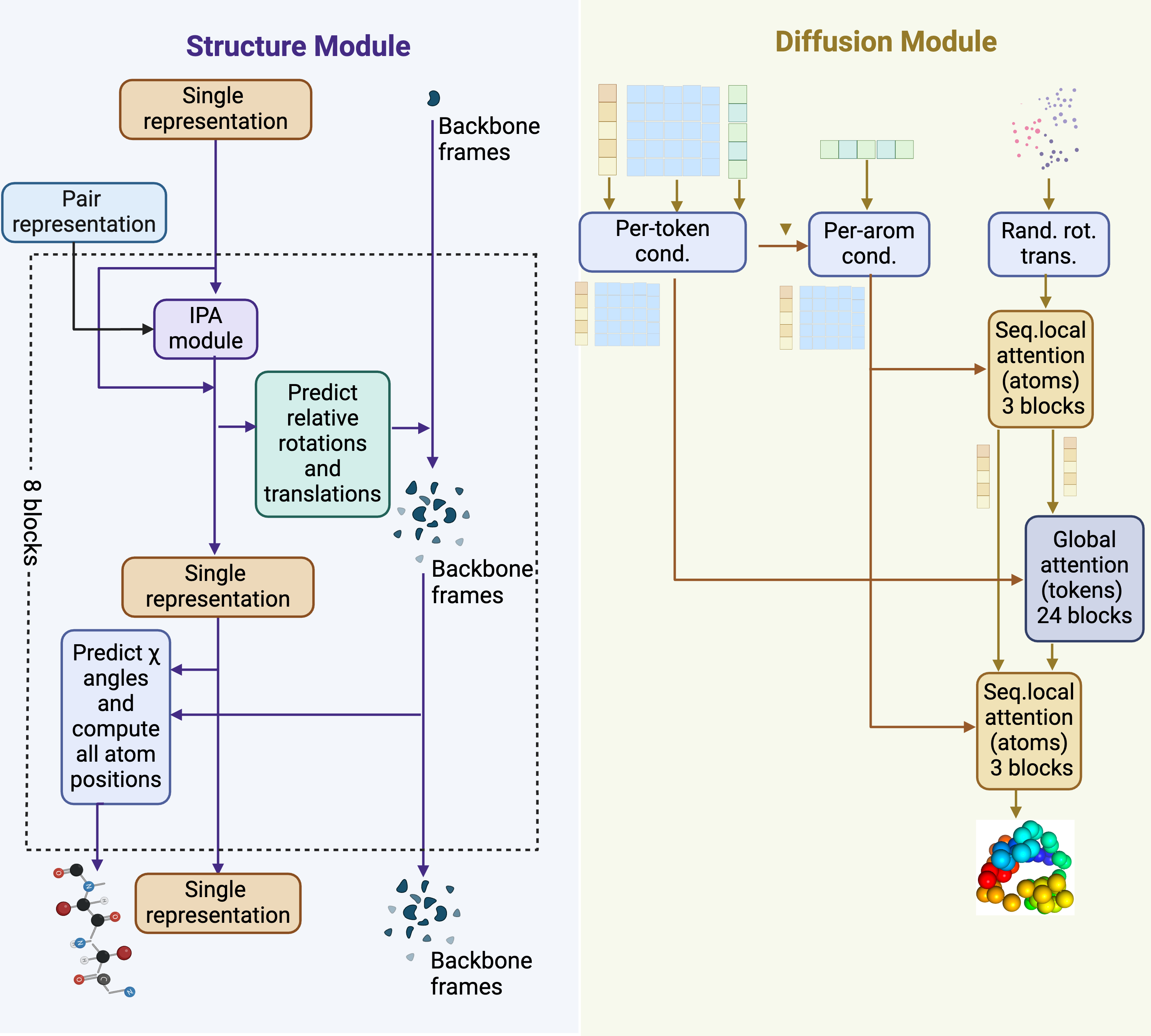}
    \end{minipage}
    \label{workflow}
    \caption{Differences between AF2 and AF3.}
    \label{alphafold}
\end{figure}

AlphaFold3 offers several improvements over its predecessor, AlphaFold2. Here, we highlight two of the most important differences and improvements that characterize AF3.
\begin{itemize}
 \item \textbf{Spatial transformation and generative prediction:} AF3 deviates from equivariant spatial transformations such as IPA used in AF2. Instead, it uses a diffusion-based approach for structure prediction.
 \item \textbf{Architectural simplifications:} The Evoformer stack used in AF2 to work with MSA and residue pairs is replaced by the Pairformer stack, which works exclusively with token pairs.
\end{itemize}
Fig. \ref{alphafold} shows details about the mentioned blocks and helps us to understand the two main differences.

Deep learning is powerful, and Alphafold has alleviated the trouble from various biologists of finding collaborators to perform protein modeling. It is indeed very nice to note that biologists feel confident and independent to try out their models. However, few outstanding issues provide scope for future research such as modeling of multi-domain proteins and modeling of unstructured regions.

\section{DiffDock}
\begin{figure}[ht]
    \centering
    \includegraphics[bb=0 0 800 300, width=12cm]{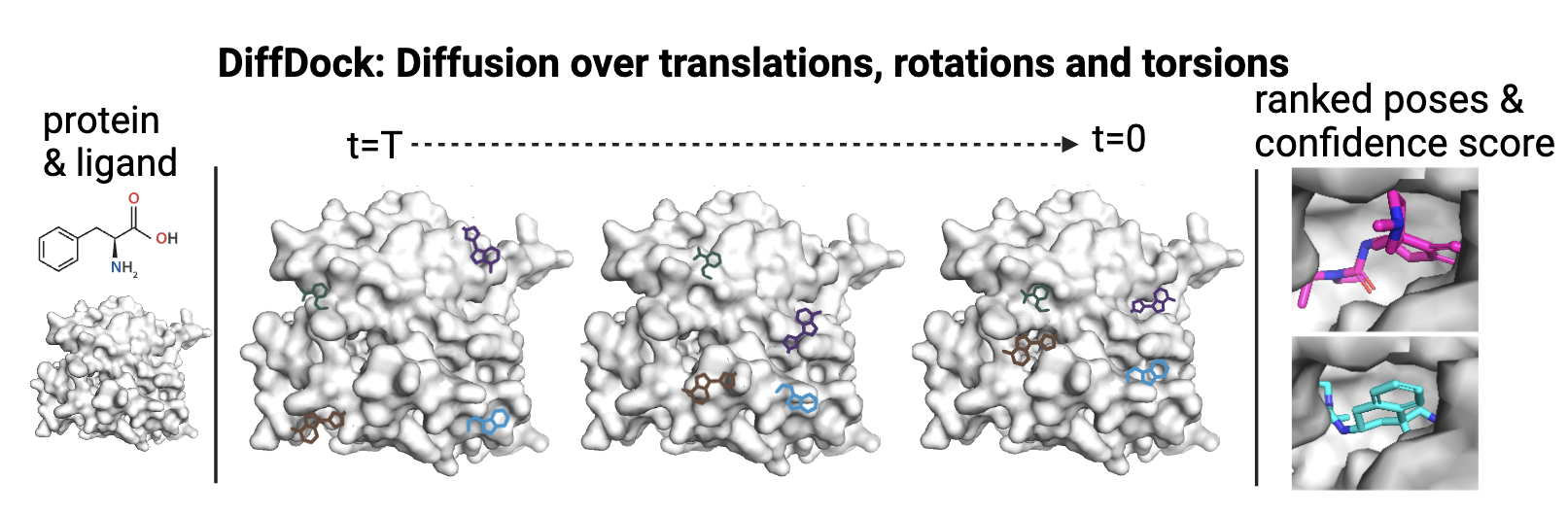}
    \caption{Visualization of DiffDock architecture.}
    \label{diffdock}
\end{figure}

DiffDock is a diffusion model tailored for protein-ligand docking that defines the diffusion process over the degrees of freedom associated with ligand poses, including ligand translations, rotations, and torsion angles (as depicted in Figure \ref{diffdock}). It achieves high selectivity and precision by employing a confidence model that forecasts confidence levels for ligand poses generated \textit{via} reverse stochastic differential equations. This model leverages the capabilities of diffusion models and adapts them specifically for the task of protein-ligand docking, presenting a novel contribution to the field.
\par

\section{Benchmarks}
\subsection{Benchmarks for protein}
\begin{figure}[ht]
    \centering
    \includegraphics[bb=0 0 500 400, width=11.5cm]{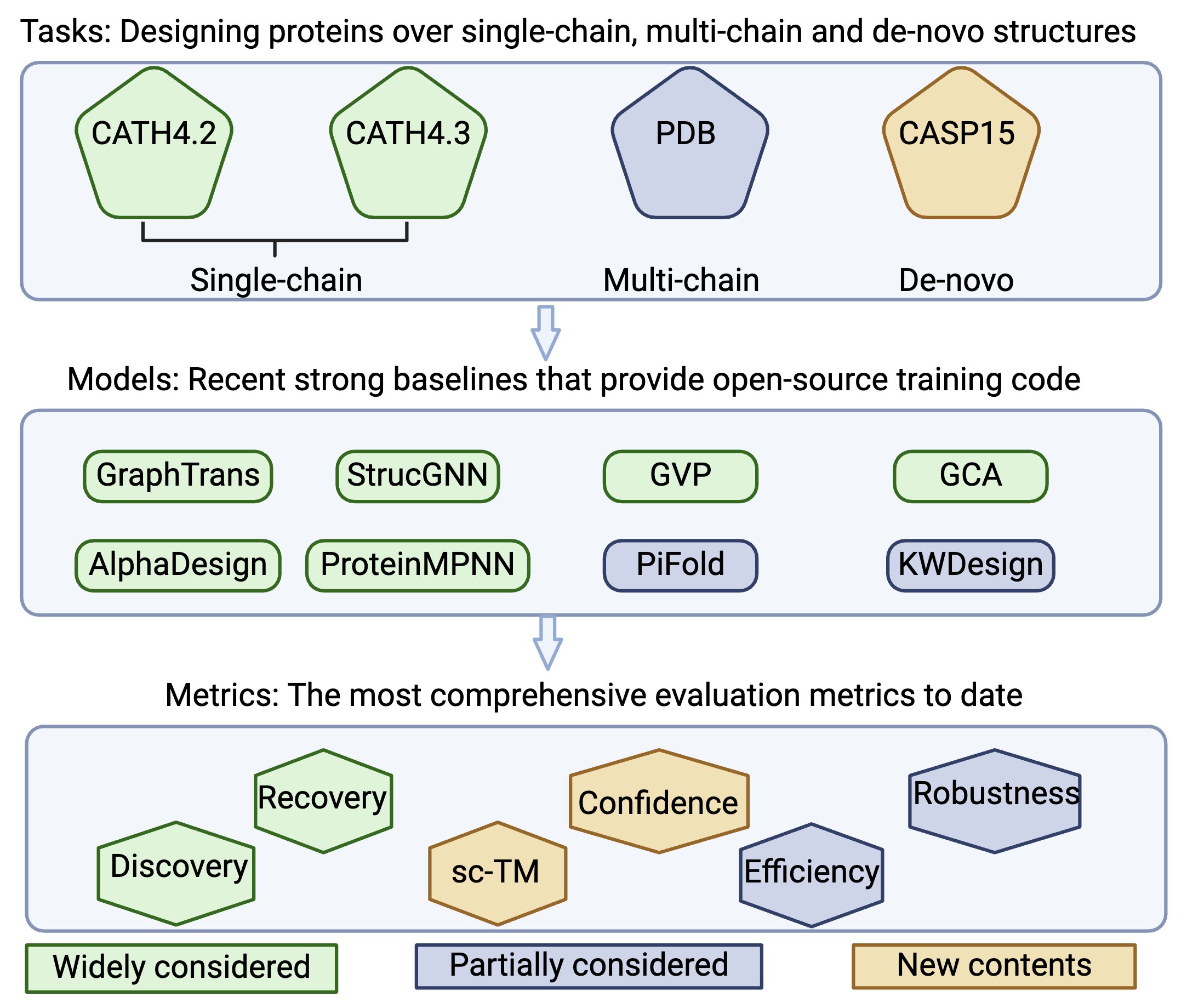}
    \caption{The framework of ProteinInvBench \cite{gao2024proteininvbench}: tasks $\Rightarrow$ models $\Rightarrow$ metrics. Green, blue, yellow: widely considered, partially considered, newly introduced contents.}
    \label{ProteinInvbench}
\end{figure}

To evaluate the performance of the models for protein backbone generation, it is crucial to establish and utilize robust benchmarks. These benchmarks not only facilitate the assessment of different generation methods, but also provide a standardized framework for comparing their strengths and limitations across various criteria. In the following, we outline several widely used benchmarks for evaluating protein backbone generation methods.
\begin{itemize}
    \item PDB-struct \cite{wang_PDBStruct} suggests that encoder-decoder methods generally outperform structure-prediction-based methods in terms of refoldability, recovery, and stability metrics. 
    \item Scaffold-Lab \cite{zheng_ScaffoldLab2024} focuses on the evaluation of unconditional generation across metrics such as designability, novelty, diversity, efficiency, and structural properties.
    \item  Melodia \cite{montalvao_Melodia2024} is a Python library with a complete set of components devised for protein structural analysis and visualization using differential geometry of three-dimensional curves and knot theory. Residue-wise confidence predicted local distance different test (pLDDT) and pairwise confidence predicted alignment error (PAE).
    \item  PINDER \cite{kovtun_PINDER2024} offers substantial advancement in the field of deep learning-based protein-protein docking and complex modeling by addressing key limitations of existing training and benchmark datasets.
    \item  ProteinInvBench \cite{gao2024proteininvbench} is a benchmark for protein design, which comprises extended protein design tasks, integrated models, and diverse evaluation metrics (see Fig. \ref{ProteinInvbench}).
\end{itemize}

\subsection{Benchmarks for molecular generation}
The goal of unconstrained molecular generation is to generate molecules that are:
\begin{itemize}
    \item Valid and unique. Validity is the percentage of valid molecules measured by RDkit, uniqueness is the percentage of unique molecules among the valid molecules.
 \item Based on a chemical distribution corresponding to the training set.
 \item Novel and diverse. Novelty is the percentage of valid molecules not found in the training set. Diversity is the opposite of recovery and is meaningless if we measure it alone. If we examine sequence diversity and structural sc-TM together, we could gain a more comprehensive understanding of the designable protein space. To expand sequence diversity, we need to allow perturbations in the conformation of the protein backbone.
\end{itemize}
\par
Continuous Automated Model Evaluation (CAMEO) \cite{https://doi.org/10.1002/prot.25431} ligand-docking evaluation, publishes weekly benchmarking results based on models collected during a 4-day prediction window and evaluates their performance. The Frachet ChemNetDistance (FCD) measure the similarity between molecules in the training set and in the test set using the embedding learned by a neural network.\par
\section{Comparison with existing reviews}
Here we list the existing reviews on diffusion models for protein design, along with their frameworks and applications, see Table \ref{tab:survey_comparison}.
\begin{table*}[ht]
    \centering
    \caption{Comparison of this review with existing surveys on Diffusion model for biomolecule generation: Frameworks and applications are enumerated}
    \label{tab:survey_comparison}
    \footnotesize
    \scalebox{0.9}{
    \begin{tabular}{l|lllll|llll}
        \toprule
        \textbf{Surveys} & \multicolumn{5}{c|}{\textbf{Frameworks}} & \multicolumn{4}{c}{\textbf{Applications}} \\ 
        \hline
                &
            \rotatebox{65}{Categorization} &  
            \rotatebox{65}{Benchmarks} & 
            \rotatebox{65}{Challenges} & 
            \rotatebox{65}{Future Works} &
            \rotatebox{65}{mathematics behind} &
            \rotatebox{65}{Protein generation} & 
            \rotatebox{65}{Peptide design} &  
            \rotatebox{65}{Molecule generation} & 
            \rotatebox{65}{Protein-ligand interaction}  \\
         \midrule
         Ours  & \cmark &  \cmark & \cmark & \cmark & \cmark & \cmark & \cmark & \cmark & \cmark \\

       \cite{norton2024sifting} & \cmark & \cmark & \xmark & \cmark & \cmark & \cmark & \xmark & \xmark & \cmark\\
       
       \cite{guoDiffusionModelsBioinformatics2023} & \cmark & \xmark & \xmark & \cmark & \xmark & \cmark & \xmark & \cmark & \cmark\\

        \cite{zhang_Survey2023} & \cmark & \cmark & \cmark  & \xmark & \cmark & \cmark & \xmark & \cmark & \cmark\\

         \bottomrule
    \end{tabular}
}
\end{table*}

\newpage
\phantomsection
\addcontentsline{toc}{section}{List of Abbreviations}
\printnomenclature 
\nomenclature{DDPM}{Denoising Diffusion Probabilistic Models}
\nomenclature{SGM}{Score-based Generative Models}
\nomenclature{DNPD}{De Novo Protein Design}
\nomenclature{VAEs}{Variational Autoencoders}
\nomenclature{GANs}{Generative Adversarial Networks}
\nomenclature{SDE}{Stochastic Differential Equation}
\nomenclature{ODE}{Ordinary Differential Equation}
\nomenclature{TaxDiff}{Taxonomic-Guided Diffusion Model}
\nomenclature{DPLM}{Diffusion Protein Language Model}
\nomenclature{RFDiffusion}{RoseTTAFold Diffusion}
\nomenclature{RFAA}{RoseTTAFold All-Atom}
\nomenclature{IPA}{Invariant Point Attention}
\nomenclature{VFN}{Vector Field Networks}
\nomenclature{FrameDipT}{FrameDiff inPainTing }
\nomenclature{TDS}{Twisted Diffusion Sampler}
\nomenclature{DFM}{Discrete Flow Model}
\nomenclature{AF2}{AlphaFold2}
\nomenclature{AF3}{AlphaFold3}
\nomenclature{PLM}{ProteinLanguage Model}
\nomenclature{AMPs}{Antimicrobial peptides}
\nomenclature{EGNN}{$E(3)$ Equivariant Graph Neural Networks}
\nomenclature{PepGLAD}{Peptide design with Geometric LAtent Diffusion}
\nomenclature{MMCD}{Multi-Modal Contrastive Diffusion model}
\nomenclature{MSE}{Mean Square Error}
\nomenclature{GDSS}{Graph Diffusion via the System of Stochastic differential equations}
\nomenclature{CDGS}{Conditional Diffusion model based on discrete Graph Structures}
\nomenclature{JODO}{joint 2D and 3D diffusion models }
\nomenclature{GNN}{Graph Neural Network}
\nomenclature{EDM}{$E(3)$ equivariant diffusion model}
\nomenclature{CGD}{context-guided diffusion }
\nomenclature{OOD}{out-of-distribution}
\nomenclature{SubDiff}{subgraph latent diffusion model}
\nomenclature{SILVR}{Selective Iterative Latent
Variable Refinement}
\nomenclature{GeoLDM}{Geometric Latent Diffusion Models}
\nomenclature{GCDM}{Geometry-Complete Diffusion Model }
\nomenclature{SBDD}{Structure-based Drug Design}
\nomenclature{rEGNN}{relaxedEGNN}
\nomenclature{MiDi}{Mixed Graph+3D Denoising Diffusion}
\nomenclature{GVP}{Geometric Vector Perception}
\nomenclature{DisCo-Diff}{Discrete-Continuous Latent Variable Diffusion Models}
\nomenclature{ETNN}{E(n)-Equivariant Topological Neural Networks}
\nomenclature{NequIP}{Neural Equivariant Interatomic Potentials}
\nomenclature{SI}{Supplementary Information}
\nomenclature{$SE(3)$}{special euclidean 3D group}
\nomenclature{$GL(n)$}{general linear group}
\nomenclature{RMSD}{Root Mean Square Deviation}
\nomenclature{pLDDT}{Predicted Local Distance Different Test}
\nomenclature{PAE}{Predicted Alignment Error}
\nomenclature{EQGAT}{Equivariant Graph Attention Networks
}


\end{document}

%% file: icml_math_commands.tex
\usepackage{amsmath,amsfonts,bm}

\def\1{\bm{1}}

\DeclareMathAlphabet{\mathsfit}{\encodingdefault}{\sfdefault}{m}{sl}
\SetMathAlphabet{\mathsfit}{bold}{\encodingdefault}{\sfdefault}{bx}{n}

